# Entropy-driven order in an array of nanomagnets

Hilal Saglam[1], Ayhan Duzgun[2], Aikaterini Kargioti[1], Nikhil Harle[3], Xiaoyu Zhang[1], Nicholas S. Bingham[1], Yuyang Lao[4], Ian Gilbert[4,5], Joseph Sklenar[4,6], Justin D. Watts,[7,8] Justin Ramberger[7], Daniel Bromley[9], Rajesh V. Chopdekar[10], Liam O'Brien[9], Chris Leighton[7], Cristiano Nisoli[**,2], and Peter Schiffer[**,1,3,4]

[1]Department of Applied Physics, Yale University, New Haven, CT 06511, USA

[2]Theoretical Division and Center for Nonlinear Studies, Los Alamos National Laboratory, Los Alamos, NM 87545, USA

[3]Department of Physics, Yale University, New Haven, CT 06511, USA

[4]Department of Physics, University of Illinois at Urbana-Champaign, Urbana, IL 61801, USA

[5]Seagate Research Group, Seagate Technology, Shakopee, MN 55379, USA

[6]Department of Physics and Astronomy, Wayne State University, Detroit, MI 48201, USA

[7]Department of Chemical Engineering and Materials Science, University of Minnesota, Minneapolis, Minnesota 55455, USA

[8]School of Physics and Astronomy, University of Minnesota, Minneapolis, Minnesota 55455, USA

[9]Department of Physics, University of Liverpool, Liverpool L69 3BX, United Kingdom

[10]Advanced Light Source, Lawrence Berkeley National Laboratory, Berkeley, CA 94720, USA

[**]Corresponding authors:  peter.schiffer@yale.edu and cristiano@lanl.gov

**Long-range ordering is typically associated with a decrease in entropy. Yet, it can also be driven by increasing entropy in certain special cases. We demonstrate that artificial spin ice arrays of single-domain nanomagnets can be designed to produce entropy-driven order. We focus on the tetris artificial spin ice structure, a highly frustrated array geometry with a zero-point Pauli entropy, which is formed by selectively creating regular vacancies on the canonical square ice lattice. We probe thermally active tetris artificial spin ice both experimentally and through simulations, measuring the magnetic moments of the individual nanomagnets. We find two-dimensional magnetic ordering in one subset of these moments, which we demonstrate to be induced by disorder (i.e., increased entropy) in another subset of the moments. In contrast with other entropy-driven systems, the discrete degrees of freedom in tetris artificial spin ice are binary and are both designable and directly observable at the microscale, and the entropy of the system is precisely calculable in simulations. This example, in which the system's interactions and ground state entropy are well-defined, expands the experimental landscape for the study of entropy-driven ordering.**

The somewhat paradoxical phenomenon of long-range ordering driven by maximizing entropy is observed in only a few systems in nature [1], such as vibrofluidized hard spheres, in which ordering maximizes the spheres' so-called free volume [2,3]. Often, ordering in one subset of degrees of freedom is driven by the possibility of increasing entropy in another subset, and thus of the total entropy. For instance, in the thin rod model of Onsager [4], rods order nematically to increase their translational entropy [5,6,7,8]. Entropy-driven ordering has been demonstrated primarily in out-of-equilibrium soft matter systems such as colloids, hard sphere suspensions, and liquid crystals, where it has importance in self-assembly [9,10] for systems of biological and technological relevance [3,5,8,11,12,13,14]. Furthermore, entropy maximization is also implicated in the formation of high entropy alloys of metallurgical importance [15].

While these effects have been studied in chemistry and in the physics of soft matter, a related yet different phenomenon in magnetic materials, so-called "order-by-disorder" [16,17], pertains instead to the interaction of spins arranged on a lattice in magnetic materials where collective excitations among magnetic moments select an ordered, rather than disordered, configuration in the ground state [18,19,20,21,22,23].

Here we report entropy-driven ordering in an artificial spin ice, a structurally ordered nanomagnet array [24,25]. Specifically, we examine entropy-driven ordering in tetris artificial spin ice (referred to as 'tetris ice' for the remainder of the paper) [26,27]. We demonstrate a different paradigm for such ordering, quite distinct from what has been previously observed. Crucially, ordering in this system has strong similarities to the soft matter systems described above, despite tetris ice being a structured nanomagnet array with no mechanical motion. Its degrees of freedom are, instead, the binary orientations of the nanoscale moments that are configured through thermalization.

Artificial spin ices can serve as models for a wide range of unusual physics unavailable in other systems, because they are lithographically designed at will. The ability to probe the magnetic degrees of freedom at the resolution of a single magnetic moment has provided the first realizations of celebrated vertex models [28] and also led to experimental demonstrations of a number of new models for collective behavior [24,25,29]. Relevant to our study, the characteristic that the individual magnetic degrees of freedom are constrained to point in one of two directions for each moment, sets tetris ice apart from the entropy-driven orderings referenced above.

The structure of tetris ice is obtained by selective removal of moments from the canonical square ice structure, as illustrated in Figure 1a and 1b. This system belongs to a category of artificial spin ices that are 'vertex-frustrated' [26,29], i.e., they are structured such that every lattice vertex cannot have its moments arranged in their local low energy configuration. As a result, the system necessarily has multiple 'unhappy vertices' that are excited out of their local vertex ground states, as well as zero-point entropy associated with the degeneracy in allocating the unhappy vertices within the lattice.

We start our discussion of the collective states of tetris ice with a description of the energy and entropy of moment configurations at low energy. As indicated in Figure 1b, the system's lowest energy manifold is composed of two different one-dimensional subsystems of alternating stripes of moments, so-called 'backbones' and 'staircases' [26,27]. In the system's ground state, the staircase (*SC*) moments contain unhappy vertices and are disordered. Moreover, the individual staircase zero-temperature disorder, and thus its correlations, are well-described by a disordered one-dimensional Ising phase [27]. In contrast, the backbone (*BB*) moments do not contain any unhappy vertices, and are ordered longitudinally, i.e., along the length of the stripes. Within a nearest neighbor coupling approximation, a given ground state configuration of the array receives no energetic advantage from being *transversely* ordered, meaning that mutual order among the different backbones is neither energetically favored nor disfavored [26].

In order to characterize ordering among the backbone moments, we use the staggered order parameter, $\Psi = (-1)^{i+j} S_{ij}^{BB}$, where $S_{ij}^{BB}$ denotes the polarization of the backbone moments, and *i* and *j* are the vertical and horizontal location indices of the moments in the underlying square ice lattice [28]. Note that $\Psi$ here is simply the standard antiferromagnetic order parameter for the ordering of square ice: the average value $\langle \Psi \rangle = \pm 1$ corresponds to the two equivalent ordered ground states of that lattice. In other words, two ordered backbones have the same $\langle \Psi \rangle$ if their moments' orientations belong to the same ground state of the underlying square lattice.

In Figure 2a, we schematically illustrate the case of neighboring backbones in the ground state with the same $\langle \Psi \rangle$. It has been proven that, in this configuration, the ground state of the staircase moments between the backbones is necessarily disordered [26]. An alternative ground state for the system has neighboring backbones alternating their values of $\langle\Psi\rangle$ between ±1, as shown in Figure 2b. In this other ground state configuration, the staircase moments between the backbones must be ordered [26]. These two alternative ground state configurations have the same energy, but the disorder in the staircases of the former gives an entropic advantage for neighboring backbones to have the same value of $\langle\Psi\rangle$. Thus, there is an entropic advantage for the entire two-dimensional system to have all backbones with either $\langle\Psi\rangle = 1$ or $\langle\Psi\rangle = -1$, implying two-dimensional order among the backbone moments. In other words, the backbones order to the same $\langle\Psi\rangle$ to gain global entropy for the system, which comes from the entropy of the disordered staircase moments. The system sacrifices the entropy that it would gain by having randomness in the value of $\langle\Psi\rangle$ among different backbones, because that entropy scales subextensively (since $\langle\Psi\rangle$ is binary, the entropy of a backbone-disordered configuration is proportional to the number of backbones in the system, which scales as the square root of the area of the array). In doing so, the system gains entropy from the staircases, which instead scales extensively, i.e., with the system size, as shown in the Supplementary Information (Section 5).

Entropy maximization implies mutual transverse ordering among backbones, but it also explains the longitudinal ordering within a single backbone. Consider a configuration of ordered backbones all with the same $\langle\Psi\rangle$, in which one backbone has a finite longitudinal domain of length $L_d$ with opposite $\langle\Psi\rangle$, as in Figure 2c. The two defects induce two ordered regions on the adjacent staircases, above and below the domain, corresponding to an approximate increase in the staircase free energy $\Delta F = K + TL_d s_{sc}$, where *K* is the energy of the domain boundaries, $T$ is the system temperature, and $s_{sc}$ is the entropy per moment of the disordered staircase (see Supplementary Materials Section 5). This entropic term in the free energy yields a constant attraction $\sim T s_{sc}$ among the two defects, suppressing the growth of the bound domain. This purely entropic interaction is crucial to explain the individual longitudinal ordering of the backbones since, as a one-dimensional system, a single backbone would not be expected to order without this attractive interaction among defects. Using similar reasoning, one can show that defects on neighboring backbones also interact entropically, favoring their alignment into two-dimensional domain walls (Figure 2d).

We now turn to experimental studies of this system. We have experimentally investigated the entropy-driven ordering in tetris ice through X-ray magnetic circular dichroism photoemission electron microscopy (XMCD-PEEM) measurements on three samples of tetris ice composed of thin permalloy ($Ni_{80}Fe_{20}$) nanoislands. The thickness (~3 nm) was chosen so that the island moments were thermally active in the measurement temperature range, i.e., thermal moment reversals occurred on the time scale of imaging. The samples (A, B, and C) had different interaction strengths between neighboring moments, associated with differences in the island size and spacing. Sample A (studied previously [27]) had the strongest interactions and sample C had the weakest interactions, based on micromagnetic calculations [30]. A representative scanning electron microscope image is shown in Figure 1a, and detailed descriptions of the samples and measurements are given in the Methods Section and the Supplementary Information (Section 1).

The XMCD-PEEM technique allows full-field imaging of moment orientations in the lattice on time scales of the order of seconds. A typical XMCD-PEEM image is shown in Figure 1c and the corresponding map of moment directions in Figure 1d. In the temperature range studied for each sample (see Section 2 in the Supplementary Information for details), the system ranged from having the moments fluctuating faster than the images could capture at the highest temperatures, to the moments being effectively frozen at the lowest temperatures. We note that this technique has been demonstrated previously to effectively thermalize the moments in artificial spin ice [31] and has been used extensively under the assumption of thermalization [24,25]. For all of the samples, the temperature dependence of the average vertex statistics is quite small, suggesting that the system thermalized at room temperature, and relaxed upon cooling to a metastable moment configuration within which the moments fluctuated without further reducing the overall system energy. This suggests that further relaxation to the ground state is limited by the complex topology of the lattice [32,33], in combination with intrinsic structural disorder associated with limitations of the lithography.

In Figures 3a and 3b, we show a schematic of the digitized moment configurations obtained from XMCD-PEEM measurements of two samples with different interaction strengths and therefore different proximity to the ground state. Figure 3a shows a moment configuration close to the ground state, demonstrating close to full two-dimensional ordering of the backbones, coexisting with disorder in the staircases. Note that the ordered configurations in the backbones correspond to those of the antiferromagnetic ground state of square ice from which the tetris ice structure is obtained, but the disordered moments on the staircases do not. This ordering is apparent in the series of so-called "type-I" vertices [2] along the backbones that correlate both along the backbones and across them, leading to visible structure in the moment orientation, i.e., the formation of near-complete loops of approximate head-to-tail flux-closure in the moment orientations, broken only by disorder on the staircases. Such structure can also be seen to a lesser extent in Figure 3b, which shows a moment configuration somewhat further from the ground state with domain walls in the backbones between regions of different $\langle \Psi \rangle$.

In Figure 3c, we show the resulting order parameter $\langle \Psi \rangle$ as a function of the average vertex energy for the different samples, noting that the different interaction strengths associated with the differences among the samples lead to different energies in the thermalized states. The average vertex energy, $E_{avg}$, is defined as $E_{avg} = \sum \varepsilon_\alpha N_\alpha / N_{total}$, where $N_\alpha$ is the number of observed vertices of type α, $\varepsilon_\alpha$ is the vertex energy, and $N_{total}$ is the total number of vertices. The vertex energies, $\varepsilon_\alpha$, were calculated using micromagnetic simulations [30] for different vertices, lattice constants, and island dimensions (see Section 3 in the Supplementary Information for details). The numbers of vertices, $N_\alpha$, were extracted from the XMCD-PEEM data. Since the temperature dependence of the vertex statistics was weak, we show data averaged over the full temperature range in which we took data (see Section 2 in the Supplementary Information for details).

Because $\langle \Psi \rangle$ in Figure 3c is measured over the entire image, its increasing value with stronger interaction energy corresponds to a transverse ordering of the moments in the backbones. In Figure 3d, we plot the fractional flipping rate of different moments in the system as a function of the interaction energy (defined as the fraction of the moment flips between successive frames that are among the backbone moments, vertical staircase moments, and horizontal staircase moments, respectively). The results show that kinetics are largely confined to the disordered staircases, especially for the largest magnitude interaction energies.

We now analyze the longitudinal and transverse correlations to quantify the two-dimensional order across the system. If $S_{ij}^{BB}$ and $S_{i'j'}^{BB}$ are two backbone moments, we define their transverse and longitudinal correlations as $C_{BB} = S_{ij}^{BB} S_{i'j'}^{BB} (-1)^{i-i'+j-j'}$, where $C_{BB}$ = +1 if the moments have the same value of $\Psi$ and $C_{BB}$ = -1 if the values of $\Psi$ are opposite. For both horizontal and vertical staircase moments, we instead define the usual ferromagnetic correlation, i.e., $C_{SC} = S_{ij}^{SC} S_{i'j'}^{SC}$. Since we are primarily interested in longer-range correlations, we consider only those horizontal staircase moments where the two adjacent moments in a step are aligned head to tail.

In Figure 4, we show measured correlations among different moment pairs for the three samples, again averaged over all temperatures; the error bars here represent standard deviations of the data collected at different temperatures. Figure 4a defines the pairs of nearest and next-nearest neighbor moments, both longitudinally and transversely, through color labels (similar definitions are in Figures 4d and 4g). Figure 4b plots longitudinal correlations of horizontal moments within the same staircases. Note that the distance dependence of the longitudinal moment correlations within the horizontal staircases does not change much among the samples, because those correlations are a property of the constrained disorder of the ground state. By contrast, there are considerable differences among the three samples in the longitudinal correlations within the backbones (Figure 4e). They become increasingly more correlated with increasing interaction strength (from C to B to A), and the correlations evolve to an almost flat $\langle C_{BB} \rangle$ = 1 value as a function of distance for sample A, as expected for the ground state.

We now consider correlations transverse to the backbones and staircases. Figure 4c shows no discernable transverse correlation among the horizontal moments of different staircases. In contrast, Figure 4f reveals considerable transverse correlations among backbone moments, with correlation values almost as large as the longitudinal case (Figure 4e). As in that case, they grow with increasing interaction strength, eventually approximating the flat $\langle C_{BB} \rangle$ = 1 value that corresponds to two-dimensional long-range order. This can be seen clearly in real space snapshots that reveal isotropic domains of various sizes (see Figure SI. 2.6 – 2.8). The contrasting complete lack of transverse order among the horizontal staircase moments is a clear indication of the separation of the backbones and the staircases in terms of their entropy – with the entropy of the backbones minimized and the entropy of the staircases (and thus of the whole system) maximized. Note that, because the staircases separate the backbones, the impressive ordering of the backbone moments is strong evidence for the entropically-mediated interactions among the backbones. Our Monte Carlo simulations, discussed below, show that a near-neighbor model, with no interaction whatsoever among backbones, replicates these experimental findings (See Figure SI. 4.2).

Completing our discussion of Figure 4, the correlations among the vertical moments in the staircases are shown in Figures 4h and 4i. We observe that the correlation among the vertical moments is almost flat in magnitude (but alternating in sign), as if they were ordered, but the value of the correlation is $|\langle C_{SC}\rangle| \sim 0.5$ for sample A, and smaller for samples B and C. These moments thus possess features of both long-range order (correlation almost constant in space) as well as disorder, in the sense that $|\langle C_{SC}\rangle|$ never approaches unity, even for near neighbors. As shown in the Supplementary Information (see Section 6), correlations among vertical staircase moments are dictated by both horizontal staircase moments, which are disordered, and also by the backbone moments surrounding the staircases, which are ordered; a value of $|\langle C_{SC}\rangle| = 0.5$ is expected in the system ground state. This sort of *half-ordering* of the vertical moments is highly unusual in magnetic systems: it is neither short-range ordering, which should approach $|\langle C_{SC}\rangle| \sim 1$ at short distances along a given lattice direction and fall sharply with increasing separation, nor a disordered state, as the absolute value of the correlation persists at nearly the same value with distance. Rather, this represents a consequence of the peculiar frustration in tetris ice.

We now discuss simulations of this system, which enhance our understanding of the experimental results. The attribution of transverse ordering to entropic effects assumes that the ordering is not arising from long-range interactions among the moments. To confirm that long-range interactions are not needed to explain the ordering, we have performed Metropolis Monte Carlo simulations within a vertex model that considers only interactions among moments that share a common vertex. This excludes interactions among the moments belonging to different backbones and staircases, and thus provides an important corroboration of the entropy-induced mechanism of the backbone ordering. Significantly, because the Monte Carlo produces a collective state that mirrors what we see in experiment, it provides a separate validation that our experimental system was well thermalized (details of the Monte Carlo results are given in the Supplementary Information, Section 4).

Figure 5a shows the entropy, the specific heat and the order parameter $\langle \Psi \rangle$ from our simulations, as a function of temperature. We note the sharp peak in the specific heat, associated with ordering of the backbones (a transition that is inaccessible in our XMCD-PEEM temperature range). The transition temperature corresponds to the energy scale of vertex interaction energies, which indicates that longer-range interactions are not driving the transition, a conclusion that is also suggested by the disorder on the staircase moments. The observation of both an ordering transition among the backbone moments, a disordered state among the staircase moments, and a residual entropy for the system, which must be associated with that disorder, shows the clear separation of the entropy among the two subsets of moments. Because the backbone moments are ordered, this simulation also provides a quantification of the entropy associated with the staircase disorder.

Figure 5b plots the corresponding temperature dependence of the moment fractional flip rates, showing how the dynamics of the system below the ordering temperature are confined to the staircase moments. Note that the fractional flip rates for both vertical and horizontal staircases are non-zero at the lowest temperatures and the vertical staircase fractional flip rate rises continuously as the temperature decreases, suggesting that those are the most active moments. This again points to the distinct behavior of the backbone and staircase moments within the tetris ice structure, despite being strongly correlated.

We also use our Monte Carlo simulations to demonstrate entropy-based ordering in a situation that cannot be easily reproduced experimentally. Specifically, we initiate the system in a ground state configuration corresponding to an order parameter of zero, i.e., $\langle \Psi \rangle = \pm 1$ on alternating backbones (see Figure SI. 4.3a). We then allow the system to evolve at a temperature of only ~0.3 $T_C$, where $T_C$ is the ordering temperature for the backbones. Our simulations show that the system spontaneously evolves through thermal fluctuations into a backbone-ordered state corresponding to uniform $\langle \Psi \rangle = 1$ (see Figure SI. 4.3b). This further supports the robust nature of the observed entropically-driven ordering among backbone moments since it can be obtained through multiple thermodynamic paths, not just through cooling from high temperature.

We now compare our entropy-driven ordering with similar phenomena in other systems. Our observed ordering in the tetris ice system is substantially different from the so-called 'order-by-disorder' in some other magnetic systems [16-23]. In those cases, fluctuations can lift the degeneracy of the ground state by selecting ordered states of lower energy excitation. In our case, however, the very ground state manifold at zero temperature 'favors' order, because this order maximizes the residual entropy that results from frustration. Indeed, configurations with ordered backbones numerically dominate the ground state manifold in the large size limit, as we demonstrate in the Supplementary Information (see Section 5).

The tetris ice system is therefore conceptually closer in nature to the entropy-based ordering seen in the structurally-disordered materials, where some degrees of freedom become ordered to enable more entropy in other degrees of freedom. A paradigmatic example is the nematic ordering of rod-shaped objects in the seminal Onsager model [1,4]. The tetris ice system similarly has two distinct and competing entropies, that of the staircase moments and that of the backbone moments. The latter is reduced to maximize the former, a mechanism which maximizes the *total* entropy, analogous to the Onsager model. An important difference, however, is that the tetris ice system is well structured around a specific geometry, with discrete degrees of freedom that are experimentally accessible. While the two entropies correspond in the Onsager model to different coordinates of the same rods, in tetris ice they refer to different positions in a lattice. The two cases are mathematically similar in that the entropy of a subset of degrees of freedom is reduced to maximize the total entropy, but the nature of the degrees of freedom are strikingly different, mechanical and continuous in the former, binary in the latter, distinctly separating the two cases.

Many groups have now established that frustration in a magnetic system can result in a residual entropy, with the spin ice pyrochlore materials providing an excellent example [34]. Our findings go considerably further, demonstrating that such residual entropy can drive robust ordering in a frustrated magnetic system. This suggests that a range of other artificial spin ice geometries could be designed to tune the balance between energetic and entropic effects in ordering of moments, a possibility that would be quite difficult to realize in other physical systems.

The observation of entropy-driven magnetic ordering in the tetris ice system also adds a new category to the types of systems that display entropy-driven ordering. While our experiments are driven purely through thermal effects, the addition of quantum fluctuations [35] will likely drive yet more exotic phenomena associated with entropic considerations. Future studies will be able to probe additional bespoke artificial spin ice structures with ground state entropy that favors other types of ordering phenomena. More generally, our results show how non-trivial forms of frustration can be used to generate unusual, even apparently paradoxical phenomena that are broadly related to other physical phenomena in disparate systems [1], and to do so in ways that enable more detailed studies of the microscopic driving behavior.

## ACKNOWLEDGEMENTS

We thank I.-A. Chioar for fruitful discussions and A. Scholl for assistance with the early XMCD-PEEM measurements. Work at Yale University and the University of Illinois at Urbana-Champaign was funded by the US Department of Energy (DOE), Office of Basic Energy Sciences, Materials Sciences and Engineering Division under Grant No. DE-SC0010778 and Grant No. DE-SC0020162 [H.S., A.K., N.H., X.Z., N.S.B., Y.L., I.G., J. S., and P.S.]. This research used resources of the Advanced Light Source, a DOE Office of Science User Facility under contract no. DE-AC02-05CH11231 [R.C.]. Work at the University of Minnesota was supported by NSF through Grant Nos. DMR-1807124 and DMR-2103711 [J.R., J.D.W., and C. L.]. Work at the University of Liverpool was supported by the UK Royal Society, Grant No. RGS\R2\180208 [D.B. and L. O.]. Work at Los Alamos National Laboratory was carried out under the auspices of the US DOE through LANL, operated by Triad National Security, LLC (Contract No. 892333218NCA000001) and financed by DOE LDRD [A.D. and C.N.].

## AUTHORS' CONTRIBUTIONS

J. Ramberger and J. D. Watts performed film depositions under the guidance of C. Leighton, and D. Bromley prepared other samples under the guidance of L. O'Brien, with H. Saglam, X. Zhang, I. Gilbert, Y. Lao, J. Sklenar, and N. S. Bingham overseeing the lithography. H. Saglam, X. Zhang, I. Gilbert, Y. Lao, J. Sklenar, N. S. Bingham and R.V. Chopdekar performed the XMCD-PEEM characterization of the thermally active samples, and H. Saglam, A. Kargioti, and N. Harle analyzed the data. H. Saglam performed micromagnetic calculations. A. Duzgun performed Monte Carlo simulations, under the guidance of C. Nisoli. C. Nisoli and P. Schiffer supervised the entire project. All authors contributed to the discussion of results and to the finalization of the manuscript.

## COMPETING INTERESTS

The authors declare no competing interests.

## DATA AVAILABILITY

Underlying data are available at the following URL: https://datadryad.org/stash/share/-sT0veB190OcBSNk3G_ZW1qMa1yjbu7zjZhs-galmV0.

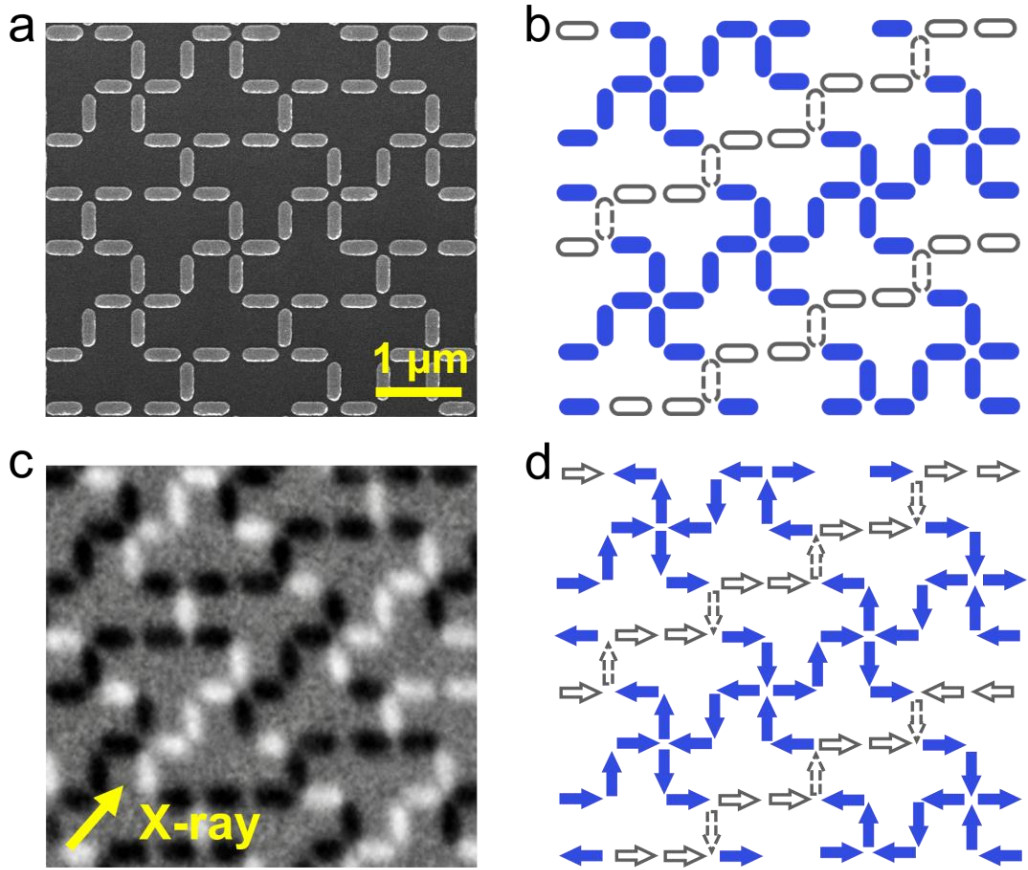

**Figure 1: Tetris Artificial Spin Ice**. **(a)** Scanning electron microscope (SEM) image of a tetris artificial spin ice (sample A). **(b)** Schematic of the tetris ice structure in which backbone nanomagnet islands are in blue and staircase islands are in grey, with the vertical moments having dashed borders and the horizontal moments having solid borders. The *longitudinal direction* is defined as the direction parallel to the stripes, approximately 26.5° degrees from the horizontal, and the *transverse direction* perpendicular to that. **(c)** X-ray magnetic circular dichroism-photoemission electron microscopy (XMCD-PEEM) image of the tetris ice lattice, where the direction of the incident X-ray beam is indicated by the yellow arrow. The islands that have a magnetization component along (opposite) the X-ray direction yield black (white) contrast (sample A at $T$ = 120 K). **(d)** Map of magnetic moment configuration corresponding to the XMCD-PEEM image in (c) with the same color scheme as (b).

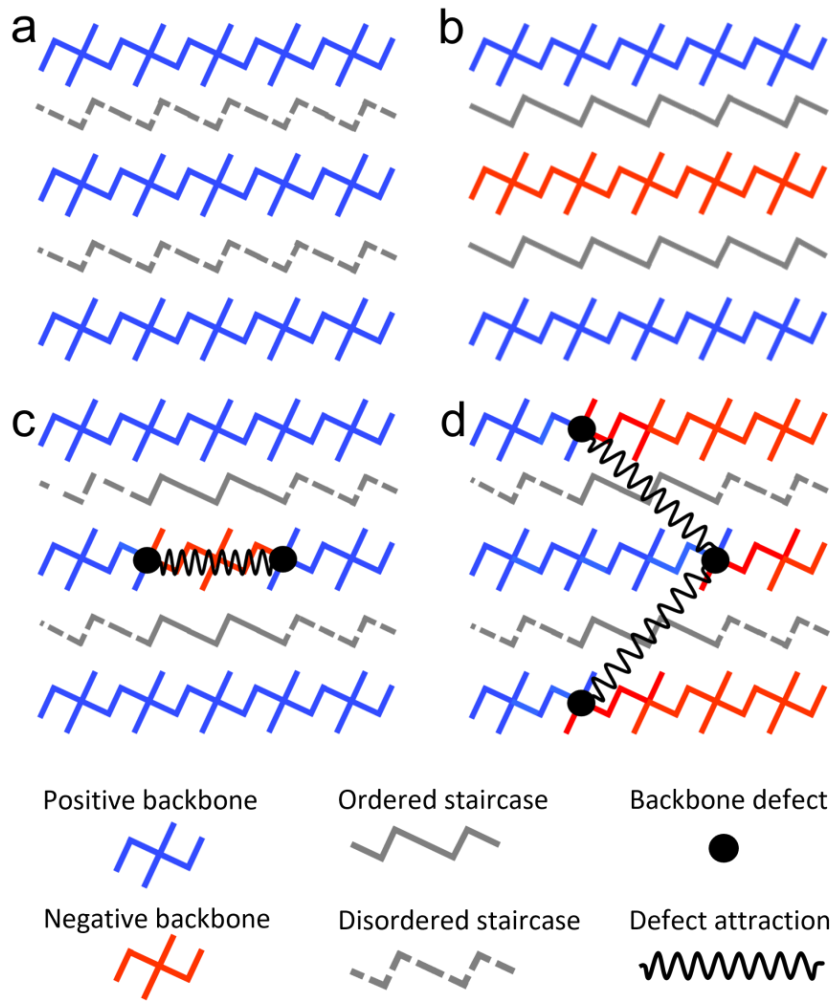

**Figure 2: Entropic Interactions in Tetris Ice**. **(a)** Schematic of a ground state configuration that leads to maximal disorder in staircases due to the transverse ordering among the backbones, all having the same $\langle \Psi \rangle$. **(b)** Schematic of a ground state configuration where backbones alternate their order parameter $\langle \Psi \rangle$, leading to the staircase moments being ordered. **(c)** Schematic of backbone defect attraction where two longitudinal defects attract each other due to the entropy cost of ordering portions of the adjacent staircases. **(d)** Schematic of backbone defect attraction across multiple backbones, where backbone defects attract each other due to the entropy cost of ordering portions of the adjacent staircases, thus favoring two-dimensional domain walls that cross multiple backbones. Moment orientations that correspond to these schematics are given in the Supplementary Information (Figure SI. 2.5)

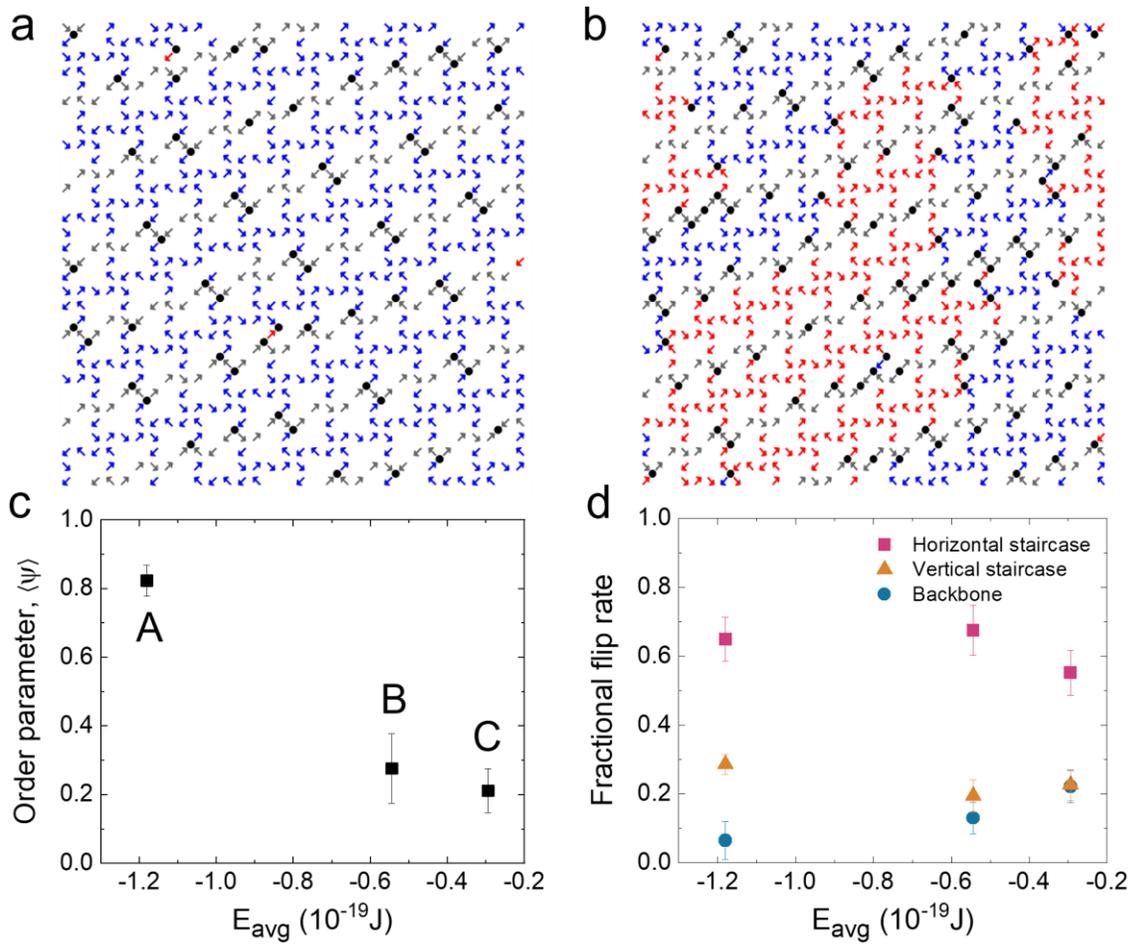

**Figure 3: Two-dimensional Ordering in Tetris Ice**. **(a)** Digitized XMCD-PEEM snapshot of tetris ice near the ground state, showing single-domain ordering of the backbones (sample A at $T$ = 120 K). **(b)** Digitized XMCD-PEEM snapshot of tetris ice above the ground state, showing backbones ordering in two-dimensional domains, while the staircases remain disordered (sample B at $T$ = 190 K). The black dots indicate unhappy vertices. **(c)** Staggered antiferromagnetic order parameter for the backbone moments plotted as a function of increasing average vertex energy from Sample A to C. **(d)** Relative probability of flipping for backbone, horizontal and vertical staircase islands, i.e., the fractional flip rate, showing that more than 80% of the kinetics is in the staircases. The error bars represent standard deviations of the data collected at different temperatures.

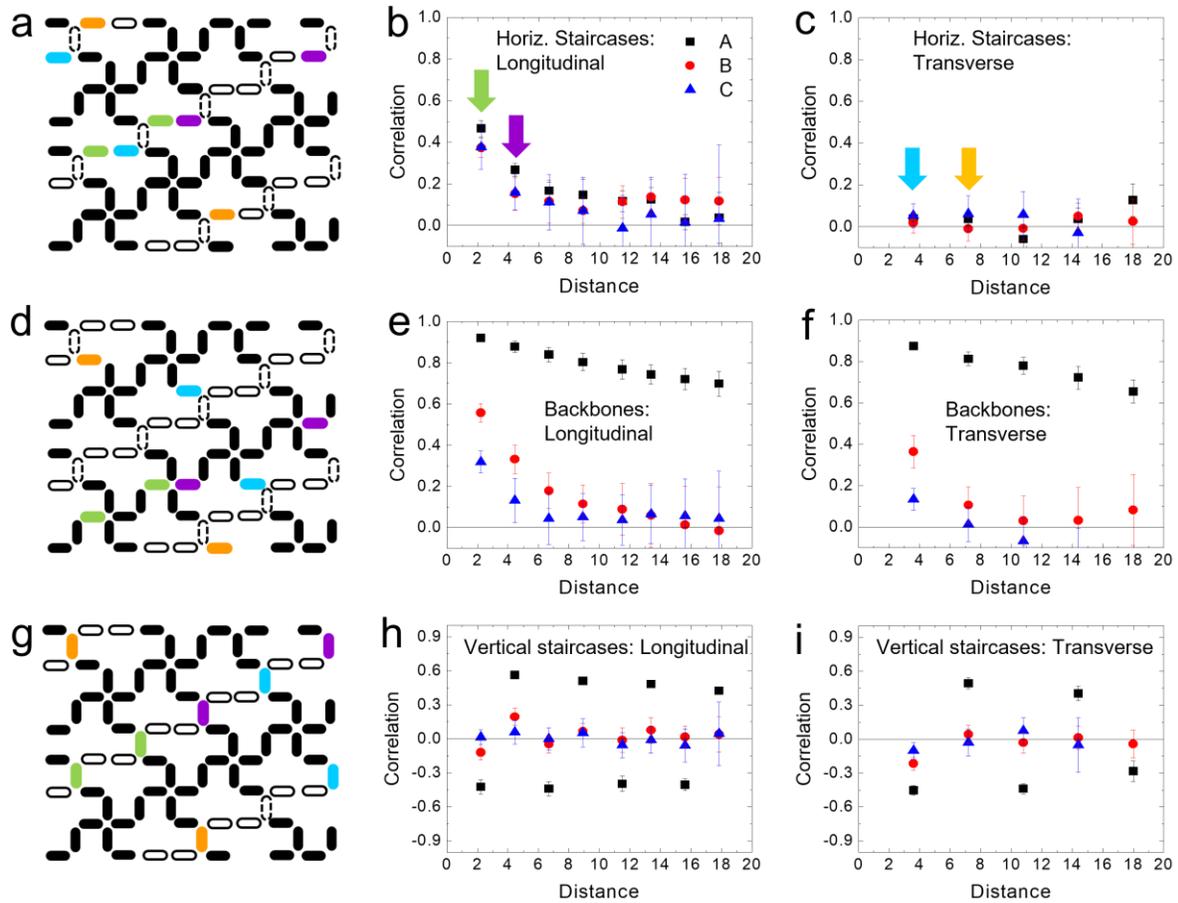

**Figure 4: Longitudinal and Transverse Moment Correlations. (a)** Schematic of the tetris ice structure highlighting the horizontal staircase islands. The two horizontal staircase islands that are each other's nearest neighbors in the longitudinal direction are colored in green. The two horizontal staircase islands that are each other's nearest neighbor in the transverse direction are colored in blue. Similarly, purple and orange indicate the next-nearest neighbors in the longitudinal and transverse directions, respectively. Note that we do not consider the pairs of horizontal moments that are within a particular stair of the staircases, since such pairs are highly correlated. **(b)** The average moment correlations as a function of distance (in the units of the lattice constant of the underlying square ice lattice structure) within horizontal staircases (longitudinal correlations) for the three samples studied. **(c)** The average moment correlations as a function of distance across the horizontal staircases (transverse correlations) for the three samples studied. **(d,e,f)** The equivalent schematics and plots as in (a,b,c) but for the backbone moments. **(g,h,i)** The equivalent schematics as in (a,b,c) but for the vertical staircase moments. The error bars represent standard deviations of the data collected at different temperatures.

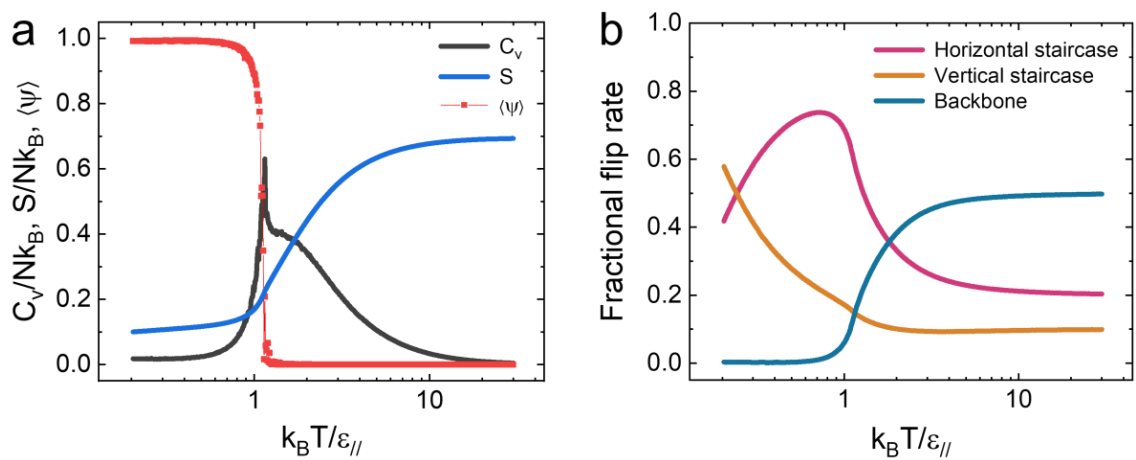

**Figure 5: Two-dimensional-Ordering in Tetris Ice - Simulation**. **(a)** Specific heat ($C_v$), entropy ($S$), and order parameter $\langle \Psi \rangle$ for the backbone moments vs. temperature computed via Monte Carlo simulations (details provided in the Supplementary Information (see Section 4)). The specific heat shows a peak corresponding to the entropy-induced ordering transition of the backbones, while the entropy shows a residual value at zero temperature that is associated with the staircases. **(b)** The temperature dependence of the moment fractional flip rates for different types of moments in the lattice.

# METHODS

Arrays of tetris artificial spin ice with various lateral dimensions and thicknesses were fabricated on silicon (Si) substrates with native oxide using electron beam lithography and lift-off as described in previous work [i,ii,iii]. The bilayer e-beam resist was spin-coated onto the substrate and exposed to the electron beam to write the desired structures. After development, permalloy ($Ni_{80}Fe_{20}$) films with varying thicknesses (2.5 nm – 3.5 nm) were deposited by ultrahigh vacuum electron beam evaporation at a rate of 0.5 Å/s. The base pressure of the system was $10^{-11} - 10^{-10}$ Torr with a deposition pressure $10^{-10} - 10^{-9}$ Torr. Subsequently, a 2 nm capping layer of Al was deposited to prevent oxidation of the permalloy. The lattice constants and island sizes were measured using scanning electron microscopy (SEM) and determined to be 602, 606, 806 nm, and 157 x 433, 157 x 433, 178 nm x 483 nm for samples A, B, C, respectively. Further details of the samples can be found in Table SI. 1.1 in the Supplementary Information.

We performed X-ray magnetic circular dichroism-photoemission electron microscopy (XMCD-PEEM) experiments on our tetris ice arrays at the PEEM-3 station at beamline 11.0.1.1 of the Advanced Light Source at Lawrence Berkley National Laboratory. Magnetic imaging was carried out at the Fe L3 edge. We conducted two XMCD-PEEM runs using different X-ray polarization sequences, exposure times, and temperature ranges. The details of the XMCD-PEEM measurements can be found in Table SI. 2.1 in the Supplementary Information.

**Methods References**

i. I. Gilbert, Y. Lao, I. Carrasquillo, L. O'Brien, J. D. Watts, M. Manno, C. Leighton, A. Scholl, C. Nisoli, and P. Schiffer, Emergent reduced dimensionality by vertex frustration in artificial spin ice, Nature Physics **12**, 162 (2016).

ii. Y. Lao, F. Caravelli, M. Sheikh, J. Sklenar, D. Gardeazabal, J. D. Watts, A. M. Albrecht, A. Scholl, K. Dahmen, C. Nisoli, and P. Schiffer, Classical topological order in the kinetics of artificial spin ice, Nature Physics **14**, 723 (2018).

iii. X. Zhang, A. Duzgun, Y. Lao, S. Subzwari, N. S. Bingham, J. Sklenar, H. Saglam, J. Ramberger, J. T. Batley, J. D. Watts, D. Bromley, R. V. Chopdekar, L. O'Brien, C. Leighton, C. Nisoli, and P. Schiffer, String Phase in an Artificial Spin Ice, Nature Communications **12**, 6514 (2021).

# Supplementary Information

# Entropy-Driven Order in an Array of Nanomagnets

Hilal Saglam[1], Ayhan Duzgun[2], Aikaterini Kargioti[1], Nikhil Harle[3], Xiaoyu Zhang[1], Nicholas S. Bingham[1], Yuyang Lao[4], Ian Gilbert[4,5], Joseph Sklenar[4,6], Justin D. Watts,[7,8] Justin Ramberger[7], Daniel Bromley[9], Rajesh V. Chopdekar[10], Liam O'Brien[9], Chris Leighton[7], Cristiano Nisoli[2], and Peter Schiffer[1,3,4]

[1]Department of Applied Physics, Yale University, New Haven, CT 06511, USA

[2]Theoretical Division and Center for Nonlinear Studies, Los Alamos National Laboratory, Los Alamos, NM 87545, USA

[3]Department of Physics, Yale University, New Haven, CT 06511, USA

[4]Department of Physics, University of Illinois at Urbana-Champaign, Urbana, IL 61801, USA

[5]Seagate Research Group, Seagate Technology, Shakopee, MN 55379, USA

[6]Department of Physics and Astronomy, Wayne State University, Detroit, MI 48201, USA

[7]Department of Chemical Engineering and Materials Science, University of Minnesota, Minneapolis, Minnesota 55455, USA

[8]School of Physics and Astronomy, University of Minnesota, Minneapolis, Minnesota 55455, USA

[9]Department of Physics, University of Liverpool, Liverpool L69 3BX, United Kingdom

[10]Advanced Light Source, Lawrence Berkeley National Laboratory, Berkeley, CA 94720, USA

## 1. Sample Fabrication

| Sample | Lattice constant (nm) | Island size (nm) | Thickness (nm) |
|---|---|---|---|
| A | 602 (10) | 157 (6) x 433 (14) | 3.5 |
| B | 606 (4) | 157 (7) x 433 (9) | 2.5 |
| C | 806 (8) | 178 (9) x 483 (13) | 2.5 |

**Table SI. 1.1:** Summary of the sample dimensions. The island sizes and lattice constants were measured using scanning electron microscopy (SEM) and averaged over 10 different islands and lattice spacings (uncertainty of approximately 10 nm, based on the standard deviation of these measurements). The thicknesses reported are subject to an approximately 10% uncertainty.

Several additional samples were also measured, but those measurements had significantly more experimental uncertainty associated with lower statistics and systemic problems with the imaging system. The resulting data were consequently more noisy, and are therefore not included here, although the qualitative behavior was consistent with the reported samples.

## 2. XMCD-PEEM Experimental Methods and Additional Data

| PEEM-XMCD run | Temperature range (K) | Temperature step (K) | Temperature cycle | Exposure time per frame (s) | Exposure sequence |
|---|---|---|---|---|---|
| A | 120 - 200 | 20 | cooling | 2.0 | 3L3R, 3 repeats |
| B | 180 - 220 | 2 | cooling & warming | 0.5 | 2L2R, 16 repeats |
| C | 180 - 220 | 2 | cooling & warming | 0.5 | 2L2R, 16 repeats |

**Table SI. 2.1:** Summary of the XMCD-PEEM experiments. The approximate numbers of imaged islands at each location were 600, 675 and 380 for Sample A, B and C, respectively. During the measurements, right (R) and left (L) circularly polarized X-rays were used for imaging and the time difference between exposures with the same and opposite polarizations of the X-rays were 0.5 s (detector/computer read-out time) and 6.5 s (detector/computer read-out time + undulator switching time), respectively.

The data presented in the main text were extracted from the XMCD-PEEM intensities of the individual tetris islands, which were converted in binary data before any quantitative analyses. In Figure SI. 2.1, we show all possible moment configurations on vertices of four islands, three islands and two islands, with coordination numbers of $z = 4$, $z = 3$ and $z = 2$, respectively. Note that the $z = 2$ coordination group can have parallel and perpendicular alignment of two-island moments. The vertex configurations represented in Figure SI. 2.1 are arranged in order of increasing energy in each coordination group, indicated by their subscripts. Note that the higher energy vertices include at least one pair of nearest neighbor moments that align head-to-head or tail-to-tail. The degeneracy of each vertex type is given by the numbers in the parentheses.

In order to calculate the flipping rate of the islands, we first found the fraction of flipped islands between two sequential images with the same polarization and divided it by the total acquisition time, which is the sum of exposure and computer read-out times of the two images. Since the vertex fractions, order parameter, and correlations show small variations as a function of temperature, we averaged all quantities over the temperature range studied for each experimental realization.

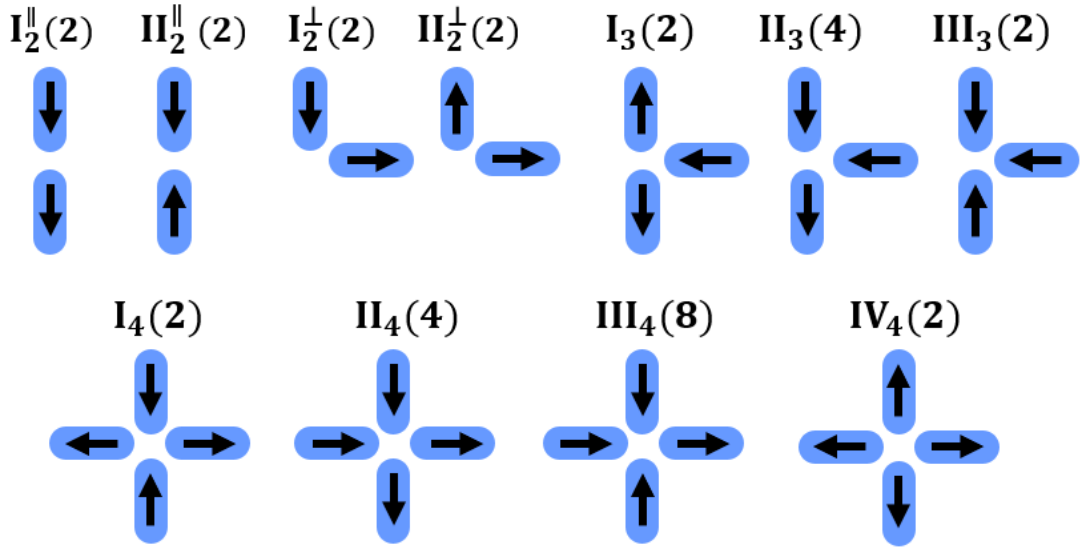

**Figure SI. 2.1:** Vertex configurations for different vertex coordination numbers $z = 2, 3, 4$, ranked in increasing energy in each coordination group. The black arrows indicate the magnetic moment direction of a given island and the numbers in the parenthesis represent the degeneracy of a corresponding vertex.

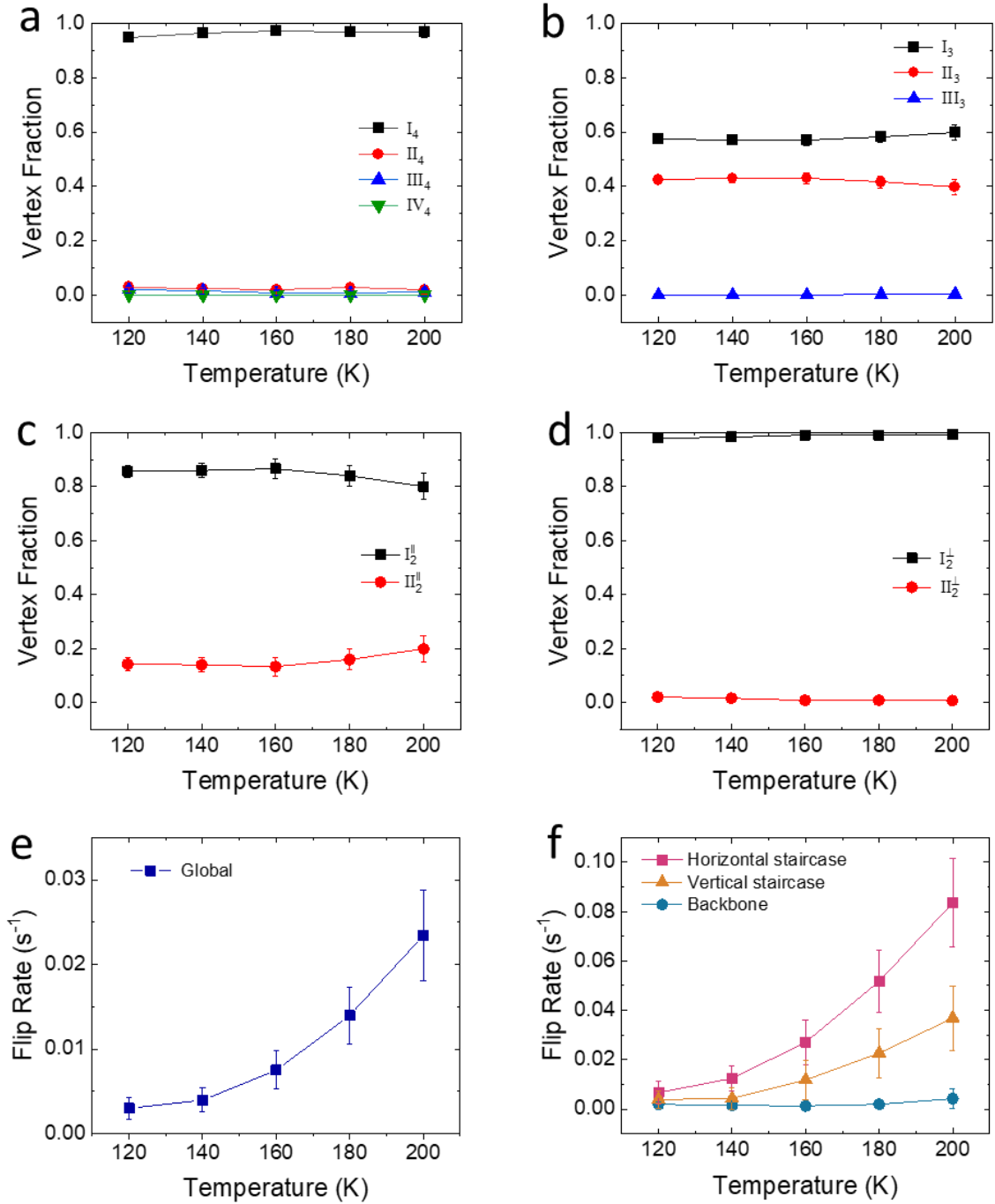

**Figure SI. 2.2:** Average vertex fractions in tetris ice as a function of temperature for coordination numbers (a) $z = 4$, (b) $z = 3$, and (c-d) $z = 2$, respectively, for Sample A with 602 nm lattice constant and 157 nm × 433 nm island dimensions. (e) Average moment flipping rate as a function of temperature for all islands. (f) Average moment flipping rate as a function of temperature for backbone, horizontal and vertical staircase islands. The error bars in all panels represent standard deviations of the data collected at 10 different locations on the same tetris array.

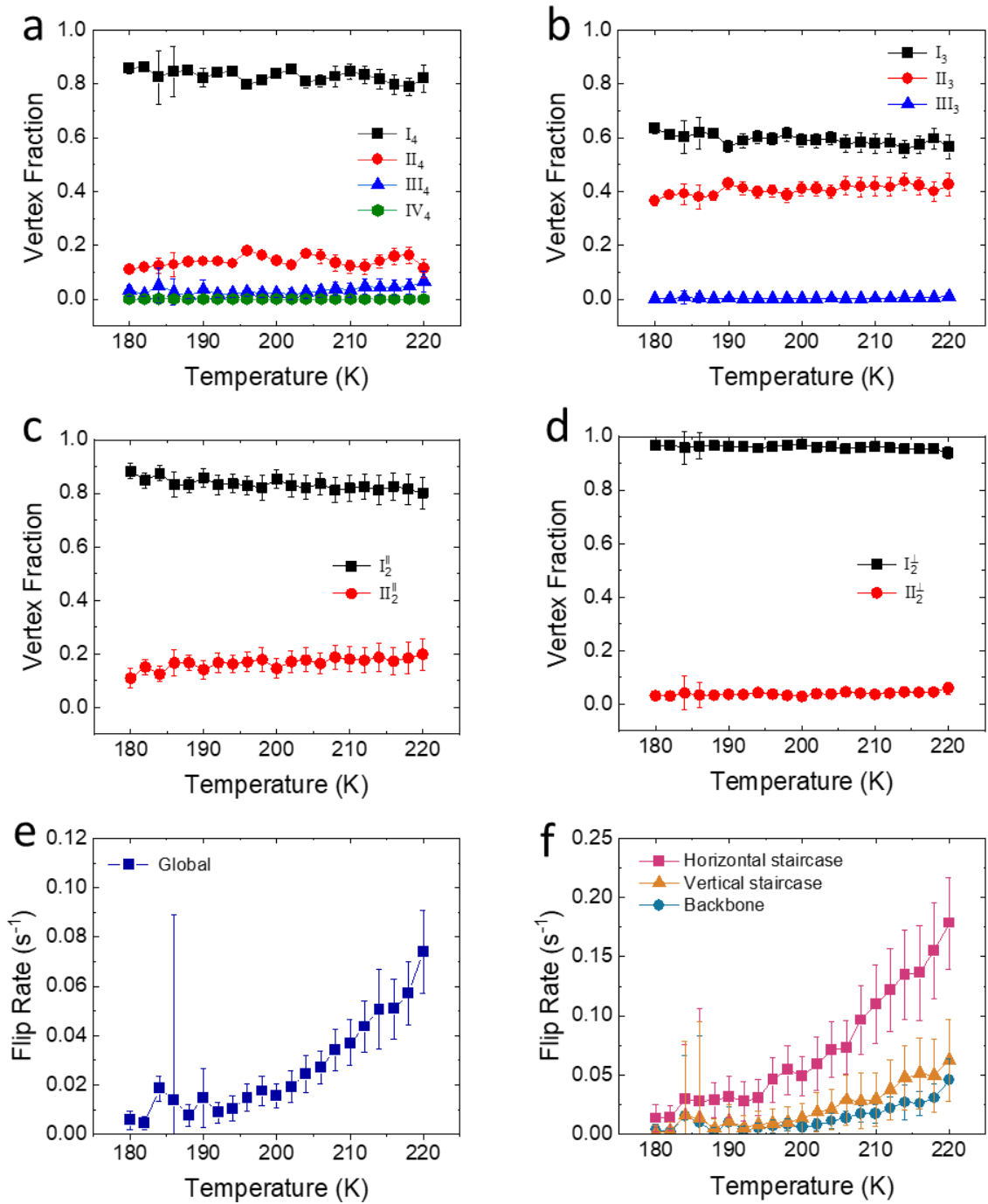

**Figure SI. 2.3:** Average vertex fractions in tetris ice as a function of temperature for coordination numbers (a) $z = 4$, (b) $z = 3$, and (c-d) $z = 2$, respectively, for Sample B with 606 nm lattice constant and 157 nm × 433 nm island dimensions. (e) Average moment flipping rate as a function of temperature for all islands. (f) Average moment flipping rate as a function of temperature for backbone, horizontal and vertical staircase islands. The error bars in all panels represent standard deviations of the data collected at 4 different locations on the same tetris array.

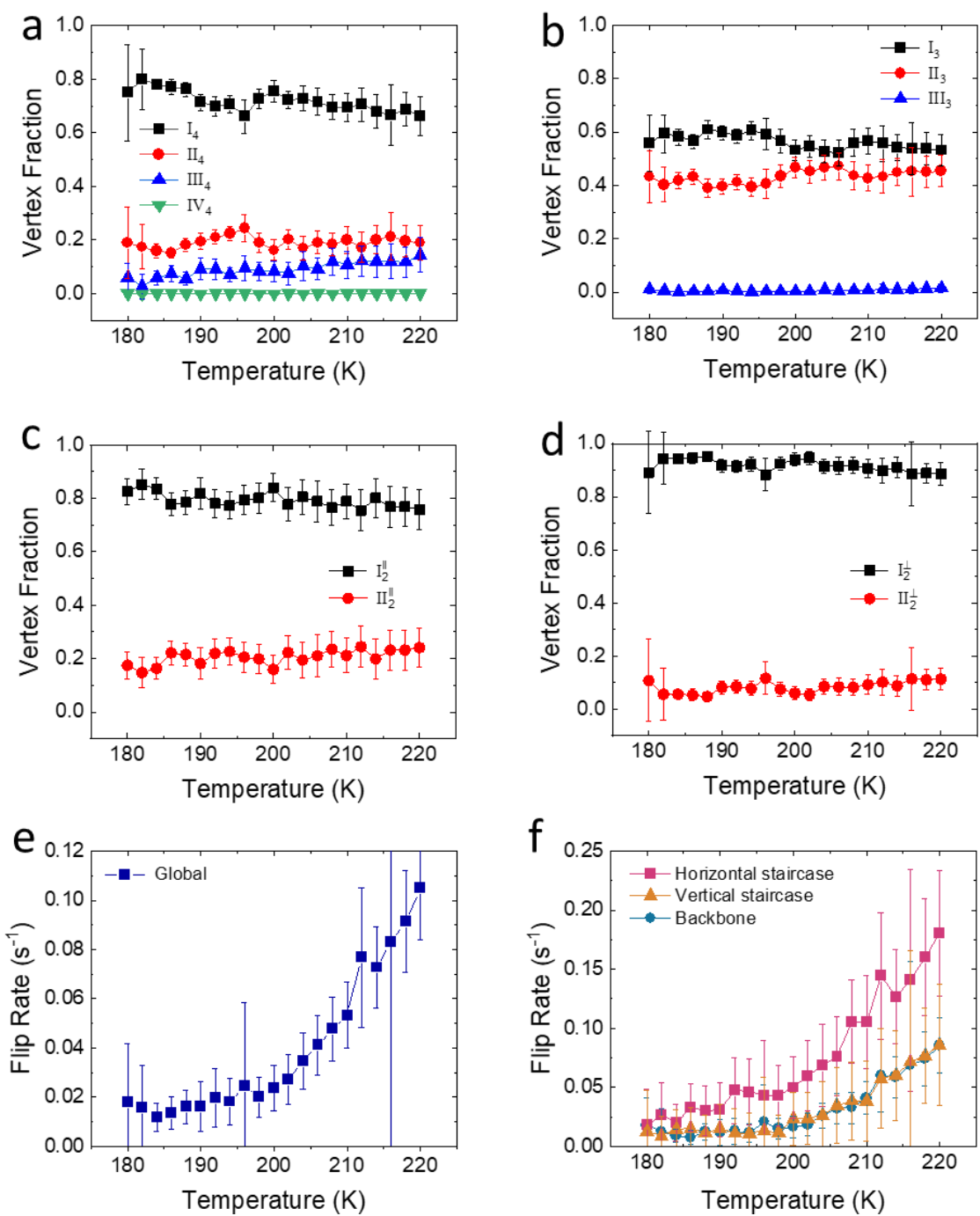

**Figure SI. 2.4:** Average vertex fractions in tetris ice as a function of temperature for coordination numbers (a) $z = 4$, (b) $z = 3$, and (c-d) $z = 2$, respectively, for Sample C with 806 nm lattice constant and 178 nm × 483 nm island dimensions. (e) Average moment flipping rate as a function of temperature for all islands. (f) Average moment flipping rate as a function of temperature for backbone, horizontal and vertical staircase islands. The error bars in all panels represent standard deviations of the data collected at 4 different locations on the same tetris array.

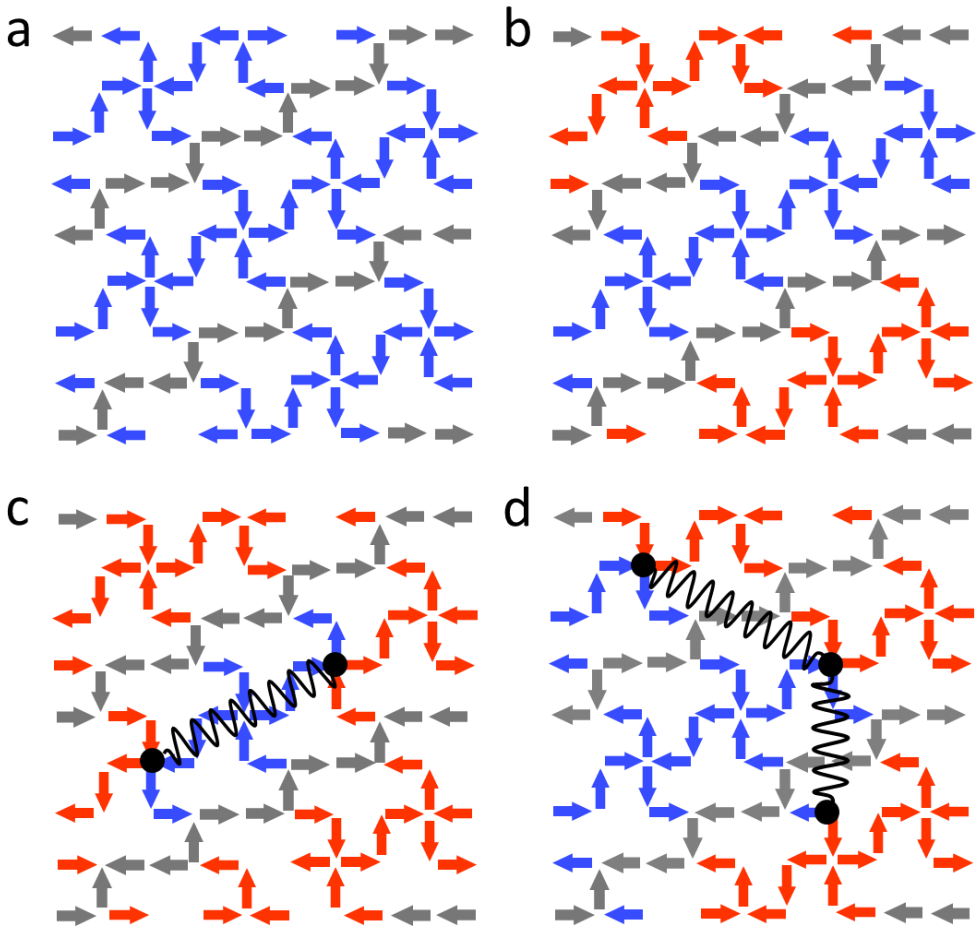

**Figure SI. 2.5:** (a) Schematic of a ground state configuration in which backbone moments are ordered both longitudinally and transversely, leading to disordered staircases. (b) Schematic of a ground state configuration where backbones alternate their order parameter only transversally, leading to ordered staircases. Note that the configurations in (a) and (b) are energetically equivalent. (c) Schematic of entropic attraction where two defects (depicted by black dots) in the same backbone attract each other (d) Schematic of entropic attraction where two defects in different backbones attract each other. Blue and red represent positive and negative backbone moments, respectively and gray indicates staircase moments. Only defects that reside on the backbones are shown. These schematics correspond to Figure 2 in the main text.

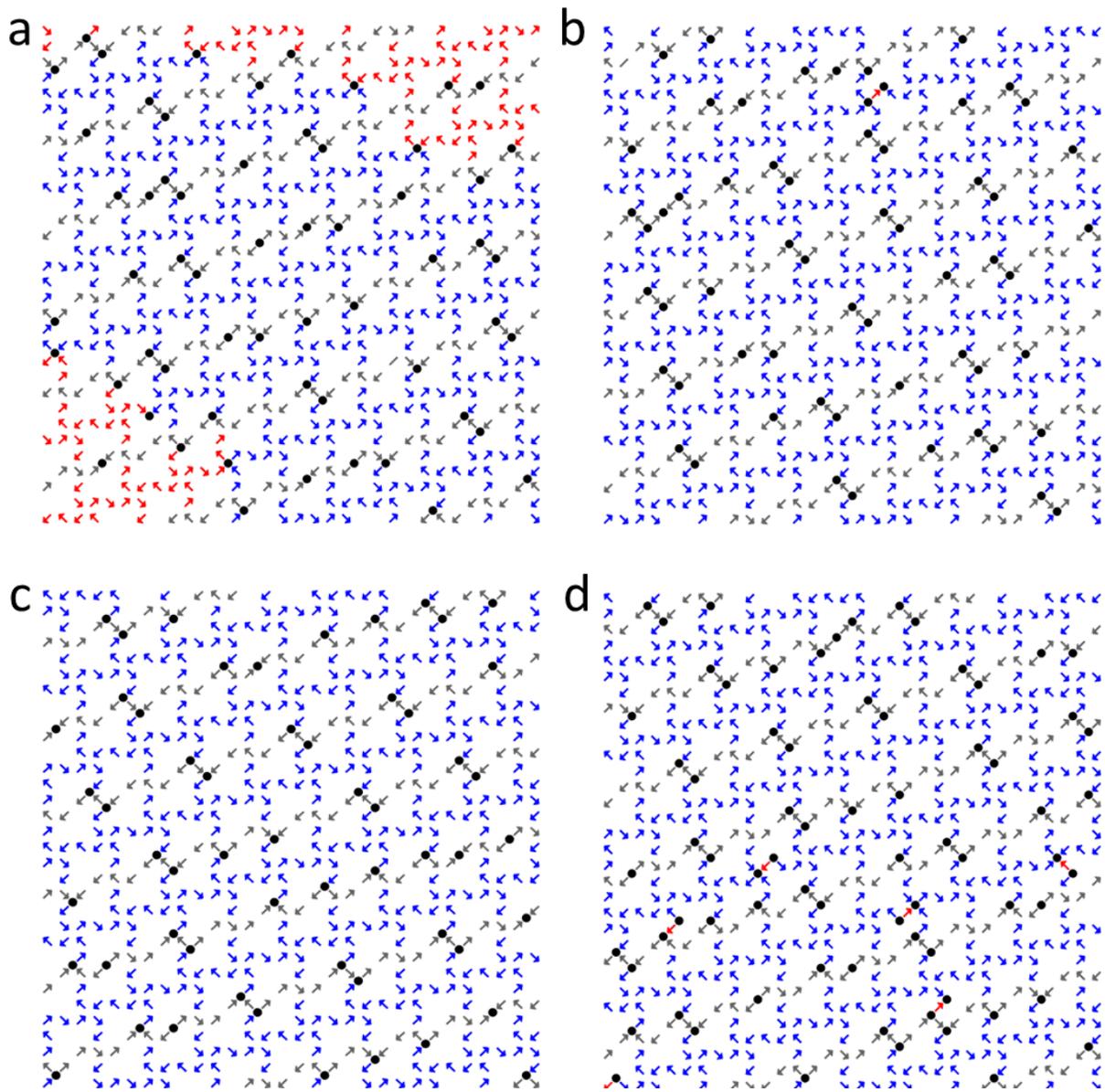

**Figure SI. 2.6:** Digitized XMCD-PEEM snapshots of tetris ice at (a) 200 K, (b) 160 K, (c) 140 K, and (d) 120 K for Sample A with 602 nm lattice constant and 157 nm × 433 nm island dimensions, showing single domain ordering of the backbones while the staircases remain disordered. Blue and red denote the backbone islands with opposite antiferromagnetic ordering, and gray represents the staircases, which are pointed according to their magnetization direction. The black dots stand for an excited state of a vertex, i.e., an unhappy vertex.

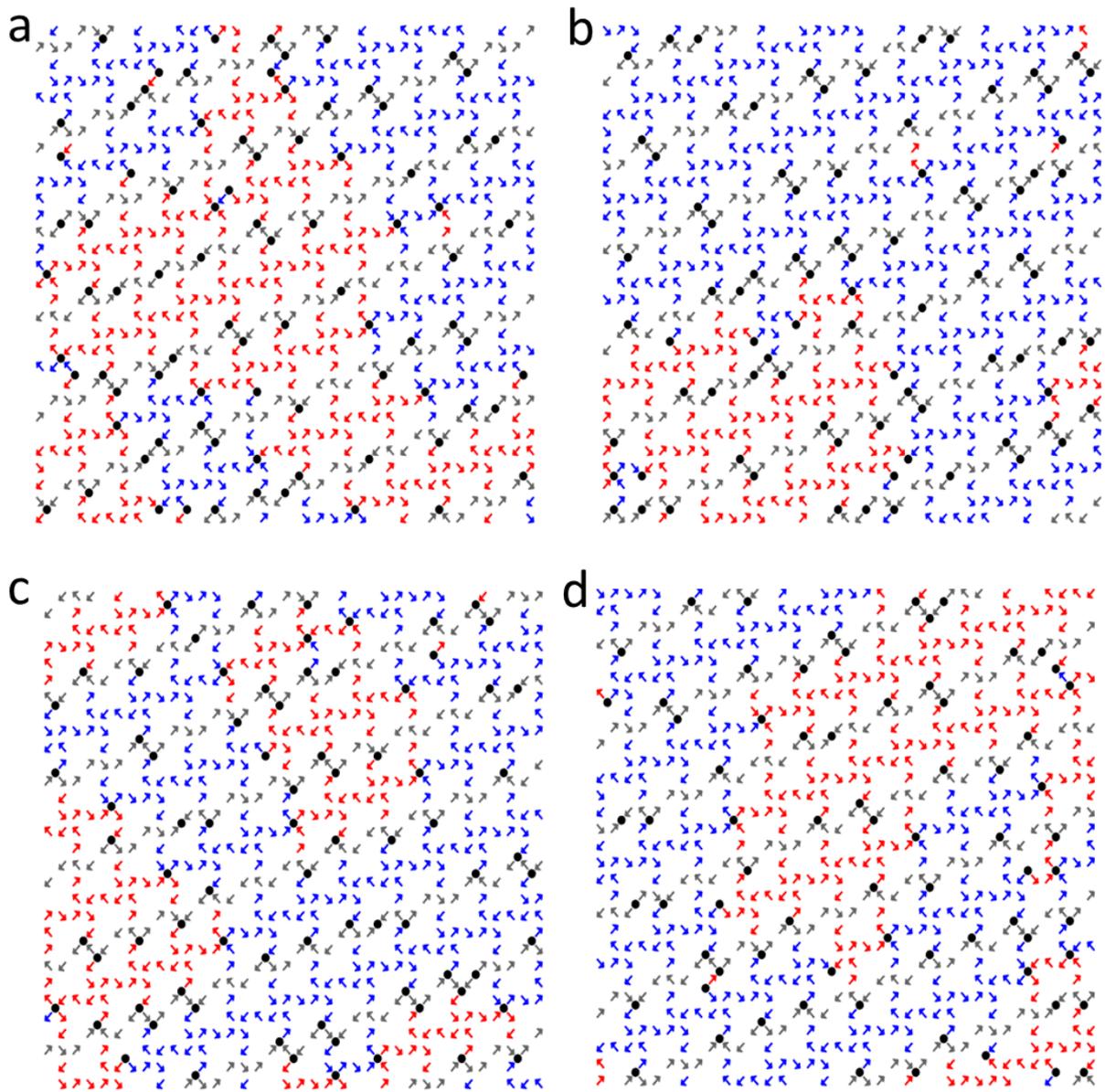

**Figure SI. 2.7:** Digitized XMCD-PEEM snapshots of tetris ice at (a) 220 K, (b) 210 K, (c) 200 K, and (d) 180 K for Sample B with 606 nm lattice constant and 157 nm × 433 nm island dimensions, demonstrating domain formation in the backbones. Blue and red denote the backbone islands with opposite antiferromagnetic ordering, and gray represents the staircases, which are pointed according to their magnetization direction. The black dots stand for an excited state of a vertex, i.e., an unhappy vertex.

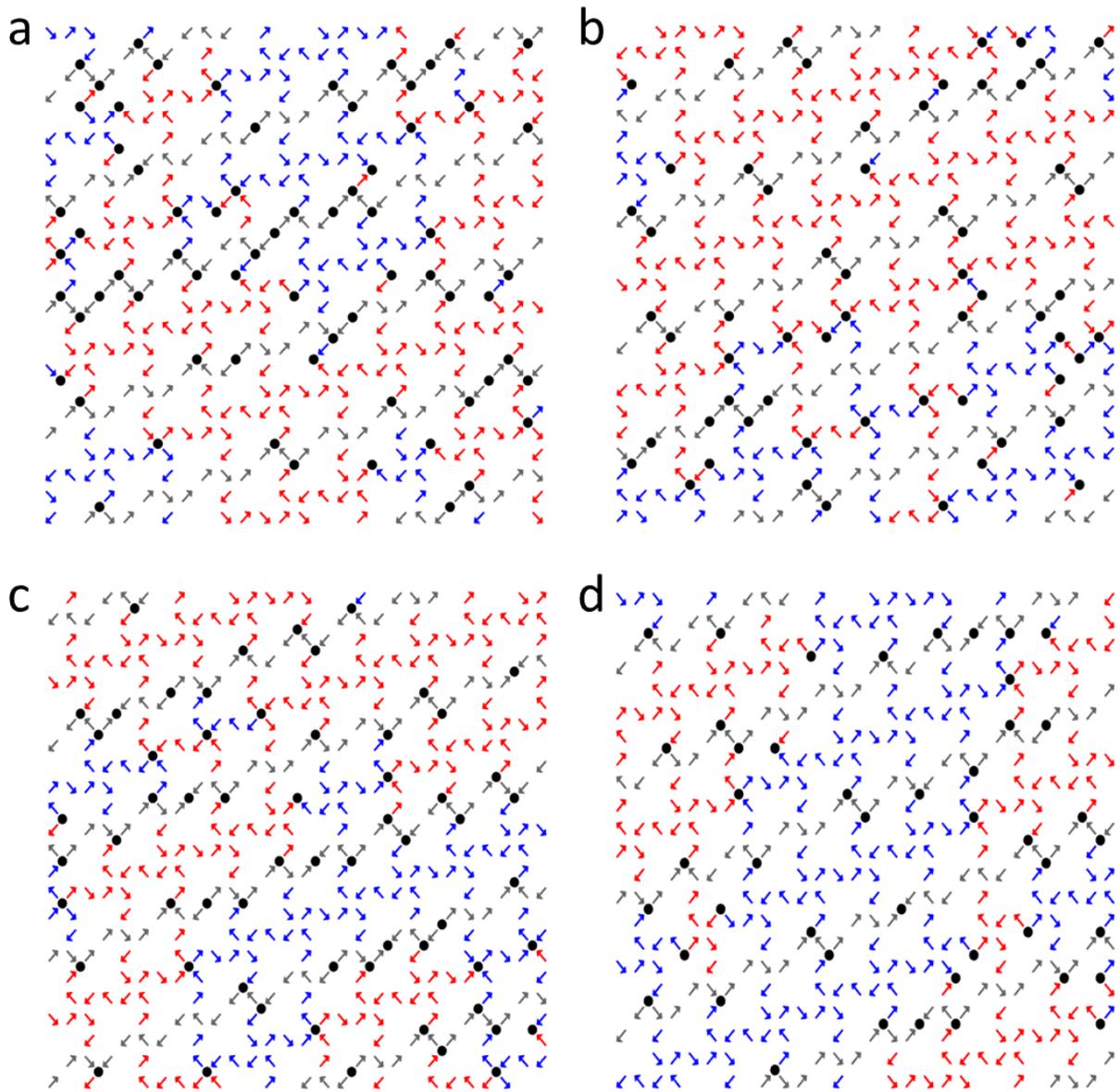

**Figure SI. 2.8:** Digitized XMCD-PEEM snapshots of tetris ice at (a) 220 K, (b) 210 K, (c) 200 K, and (d) 180 K for Sample C with 806 nm lattice constant and 178 nm × 483 nm, showing that the backbones form smaller domains due to the decreasing interaction energy. Blue and red denote the backbone islands with opposite antiferromagnetic ordering, and gray represents the staircases, which are pointed according to their magnetization direction. The black dots stand for an excited state of a vertex, i.e., an unhappy vertex.

## 3. Micromagnetic Simulations

Micromagnetic simulations were performed with the *MuMax* software package [1]. The following standard parameters for permalloy were adopted; saturation magnetization $\mu_0 M_s$ = 860 A/m and exchange constant $A_{ex}$ = 13x10$^{-12}$ J/m. The sample dimensions used in the simulations were extracted from SEM images (see Table SI.1.1).

We define the average vertex energy of the system $E_{avg}$ by the following expression:

$$E_{avg} = \sum \varepsilon_\alpha N_\alpha / N_{total} = (E_{I_4} N_{I_4} + E_{II_4} N_{II_4} + E_{III_4} N_{III_4} + E_{IV_4} N_{IV_4} + E_{I_3} N_{I_3} + E_{II_3} N_{II_3} + E_{III_3} N_{III_3}$$
$$+ E_{I_2^\parallel} N_{I_2^\parallel} + E_{II_2^\parallel} N_{II_2^\parallel} + E_{I_2^\perp} N_{I_2^\perp} + E_{II_2^\perp} N_{II_2^\perp}) / N_{total}, \quad (3.1)$$

where the subscripts correspond to the energy $E$ and the number of vertices $N$ for different vertex configurations and $N_{total}$ is the number of all types of vertices. In order to calculate $E_{avg}$, we first obtained the coupling energies of the parallel and perpendicular $z = 2$ islands by

$$\epsilon_\parallel = [\epsilon_{II2}(\parallel) - \epsilon_{I2}(\parallel)]/2, \quad (3.2)$$

$$\epsilon_\perp = [\epsilon_{II2}(\perp) - \epsilon_{I2}(\perp)]/2, \quad (3.3)$$

where $\epsilon_{II2}(\parallel)$, $\epsilon_{I2}(\parallel)$, $\epsilon_{II2}(\perp)$, and $\epsilon_{I2}(\perp)$ are the energies of given moment configurations, shown in Figure SI. 2.1. Then, the vertex energies for each configuration were defined as a superposition of $\epsilon_\parallel$ and $\epsilon_\perp$, e.g., $E_{I_4} = -4\epsilon_\perp + 2\epsilon_\parallel$.

| Sample | $\epsilon_{I2}(\parallel)$ | $\epsilon_{II2}(\parallel)$ | $\epsilon_{I2}(\perp)$ | $\epsilon_{II2}(\perp)$ |
|---|---|---|---|---|
| A | 2.142 | 2.253 | 2.06 | 2.268 |
| B | 1.199 | 1.254 | 1.164 | 1.264 |
| C | 1.403 | 1.435 | 1.382 | 1.441 |

**Table SI. 3.1:** Simulated energies for ground and excited states of the vertices with coordination number $z=2$, given in the unit of 10$^{-18}$ J. Sample parameters (lattice constant, island size and thickness) used in the simulations were the same as experimentally obtained values reported in Table SI. 1.1.

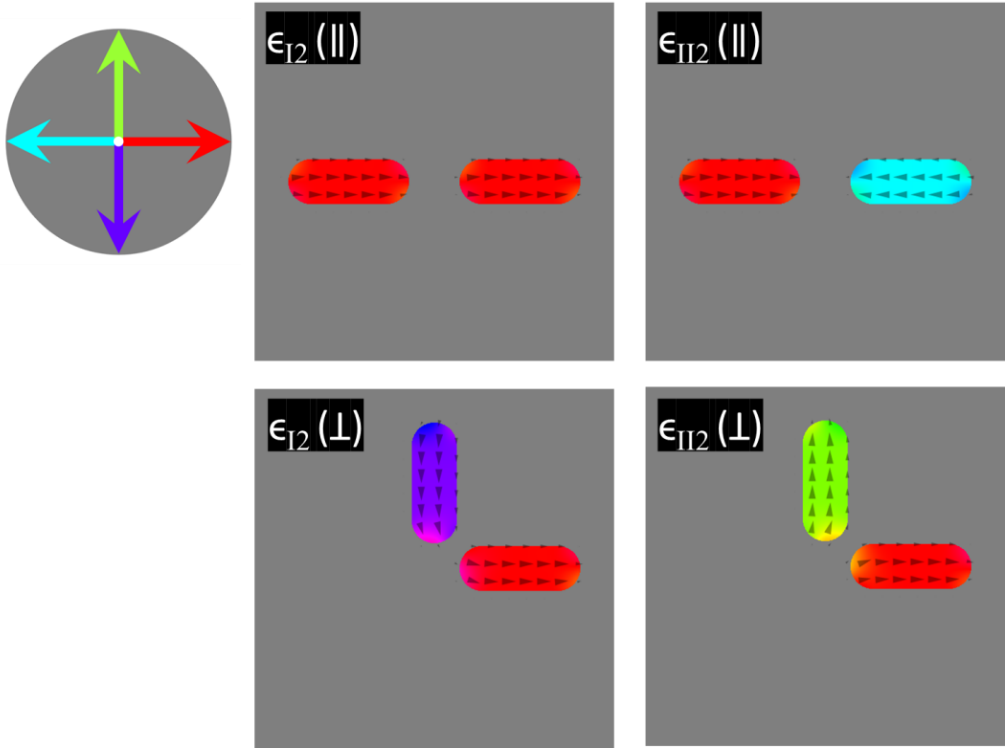

**Figure SI. 3.1:** Representative micromagnetic results for Sample A with 602 nm lattice constant and 157 nm × 433 nm island dimensions. The arrows represent the local magnetization directions within a given island.

## 4. Monte Carlo Simulations

**Model:**

We have run Monte Carlo simulations of large systems using GPU-accelerated code run on Nvidia Quadro RTX-8000 graphics processing units. The moment configurations in the system were updated with trials of single spin flips, which are either accepted or rejected according to the Metropolis algorithm. The energy of the system was calculated assuming a vertex model, in which the range of interaction between pairs of spins is limited to the nearest neighbors that meet at the common vertex. At each vertex, the interaction between a pair of spins is $\pm\epsilon_\parallel$ for parallel spins and $\pm\epsilon_\perp$ for perpendicular spins. In both cases, the sign is negative for spins that have head-to-tail alignment and positive for the opposite alignment. The ratio $\epsilon_\perp/\epsilon_\parallel$ was estimated from micromagnetic simulations and set to 1.8 for all simulations presented in this work. The unit of energy in the simulations were set such that $\epsilon_\parallel = 1$ and thus $\epsilon_\perp = 1.8$. In experimental realizations, these values vary depending on the lattice spacing as well as the shape and the magnetic properties of the individual islands.

**Simulations shown in Figure 5 of the main text:**

Phase behavior was obtained from simulations where the system was initiated at high temperature and slowly cooled to zero. The system of 128 x 128 square lattice points which hosts 20480 spins was updated $10^8$ times with 20480 flip trials at each time step. Therefore, the total simulation time was $10^8$ trials/spin. The initial temperature of $k_B T/\epsilon_\parallel = 30$ was exponentially reduced, at a cooling rate which roughly corresponds to setting the temperature to 99% of the previous value at each of the 500 discretized steps. At each temperature step, $10^8/500 = 2 \times 10^5$ flip trails per spin were performed and using the last $1.6 \times 10^5$ trials per spin, the heat capacity and the entropy were calculated from the thermal fluctuations in energy. The heat capacity $C_v$ per spin was calculated using the equation

$$\frac{C_v}{k_B} = \frac{1}{(k_B T)^2} \frac{\sigma^2(E_N)}{N} \tag{4.1}$$

where $E_N$ is the total energy of the system which has $N$ spins, and $\sigma$ is the standard deviation of $E_N$. The entropy per spin is calculated by

$$\frac{S(T)}{k_B} = \ln 2 + \int_\infty^T \frac{C_v}{k_B} \frac{dT}{T}. \tag{4.2}$$

Here, for the highly disordered initial state at $k_B T/\epsilon_\parallel = 30$, we assume that the entropy per spin approximately has the infinite temperature value of $\ln 2$. Both the curve for the order parameter $\langle \Psi \rangle$, and the peak in the curve for the heat capacity $C_v$, reflect that the ordering temperature, $T_C$ for the backbones is around $k_B T/\epsilon_\parallel = 1.1$. The collective moment state of the backbones above this temperature includes excitations away from the ground state ordering, corresponding the excited vertices in the underlying square ice structure.

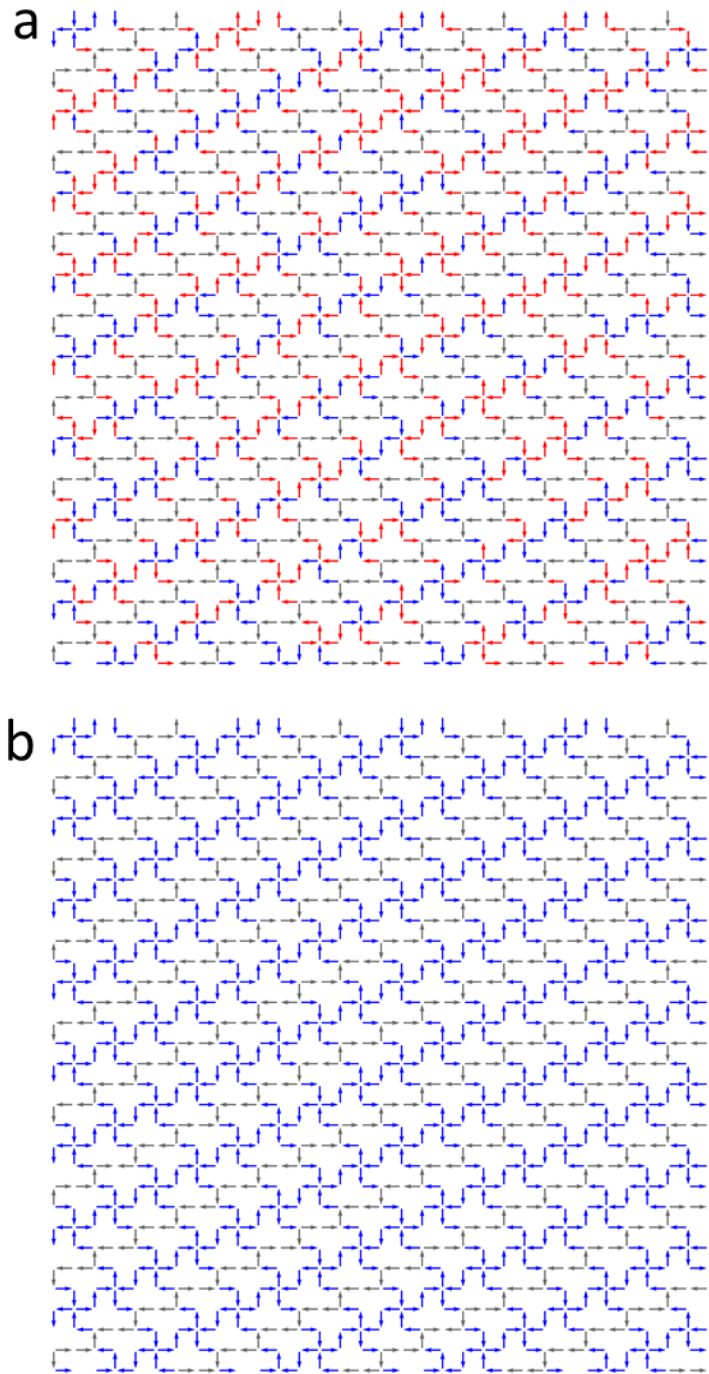

**Figure SI. 4.1:** (a) Snapshot from the initial disordered state at high temperature of the simulations shown in Figure 5 of the main text. (b) Snapshot from the final ordered state at zero temperature of the simulations shown in Figure 5 of the main text. The color convention is identical to the experimental snapshots, i.e., blue and red arrows represent the backbone islands with opposite antiferromagnetic ordering, and gray arrows represent the staircase moments.

**Correlations:**

For comparison with the experimental data, we also looked at the correlations between spins within and across backbones and staircases using the data from the same simulations described above (simulations shown in Figure 5 of the main text). The method of calculation was the same as for the experiments, which were shown in Figure 4: staggered antiferromagnetic correlation for the backbones and ferromagnetic correlation for the staircases. In Figure SI. 4.2, we show the correlations calculated at five different values of temperatures, above and below the ordering temperature, $T_C$ of $k_B T/\epsilon_\parallel \approx 1.1$. In Figure SI. 4.2c and 4.2d, we observe that the backbones are ordered both longitudinally and transversely while Figure SI. 4.1a and 4.1b show that horizontal staircases are only ordered longitudinally with short correlation length as described in [2]. Strikingly, the correlations between vertical staircases are in great agreement with the experimental results (Figure SI. 4.2e and 4.2f) and in the simulations we observe that vertical staircases are disordered above $T_C$ and partially ordered below $T_C$. These simulation results demonstrate that long-range interactions are not required for the transverse ordering of backbones, because the vertex model of spin interactions in the simulations includes only nearest neighbors.

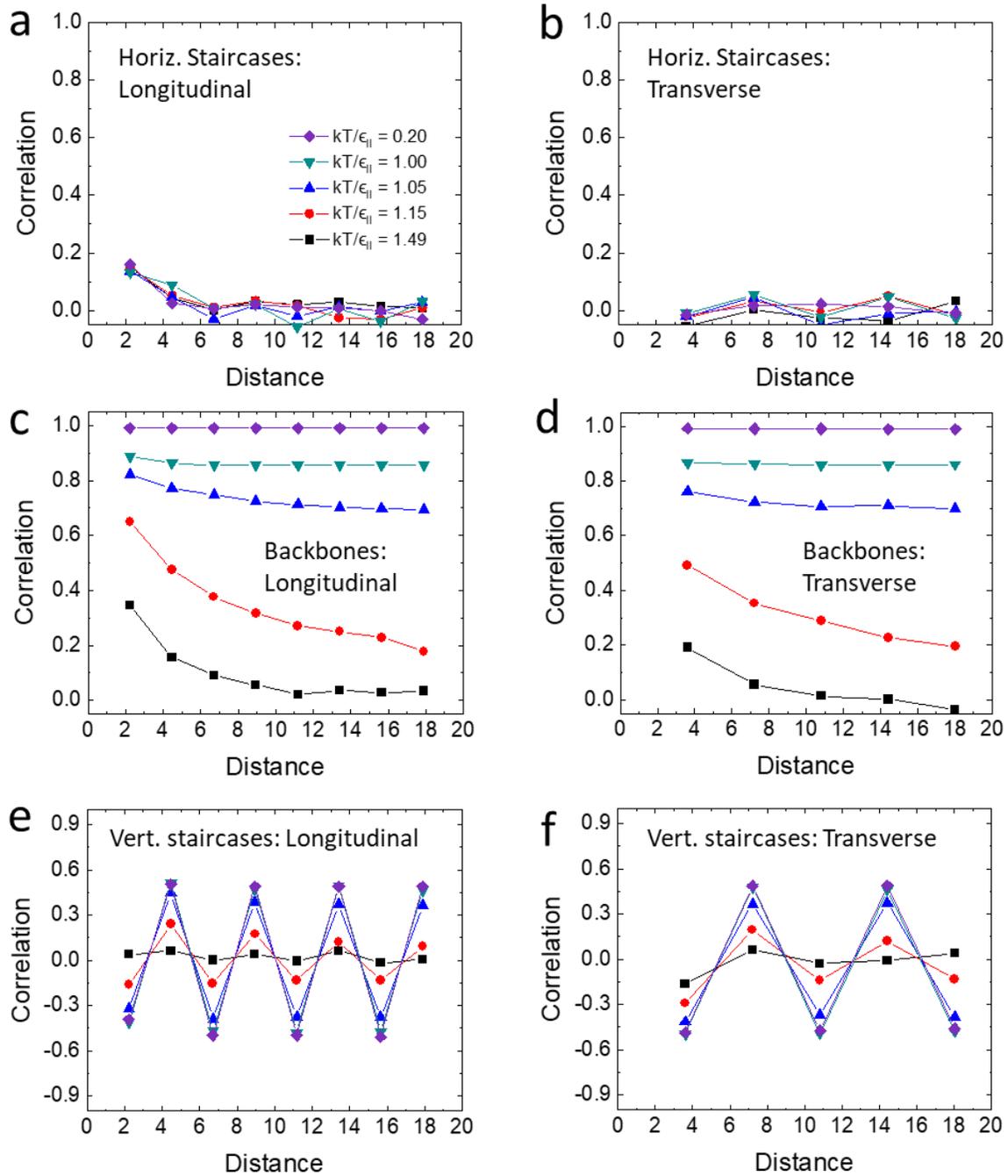

**Figure SI. 4.2:** Simulated correlations (a) within and (b) across the horizontal staircases, (c) within and (d) across the backbones, and (e) within and (f) across the vertical staircases as a function of real distance in the unit of lattice constant for different temperatures.

**Low temperature evolution from backbone-disordered ground state to backbone-ordered ground state:**

Tetris ice has degenerate ground states. Backbones with antiferromagnetic order of either sign, namely $\langle \Psi \rangle = +1$ or $\langle \Psi \rangle = -1$, can be arranged in any configuration and the system has the same ground state energy. They can be all +1 or all -1 or a mixed combination because a minimum energy configuration is accessible for the staircase spins in any combination with the following constraint: a staircase between two adjacent backbones of the same sign must be disordered and a staircase between two backbones of opposite sign must be ordered, as also explained in the main text.

We conducted a separate Monte Carlo simulation to explore the behavior of the system when it is started from a low entropy ground state. Specifically, we initiated the system in a state (Figure SI. 4.3a), where the backbones had alternating signs of $\langle \Psi \rangle$ and the staircases were ordered with head-to-tail alignment. Throughout the simulation, the temperature was kept constant at $k_B T/\epsilon_\parallel = 0.4$, which is ~0.3 $T_C$. The simulation time was 6 x 10$^8$ time steps where 1 step included as many flip trials as the number of the spins in the system. The system size was 128 x 128 lattice points with 20480 spins and all other parameters are the same as for the rest of the simulations.

The system was initiated in the alternating backbone ground state as described above, and it then was allowed to evolve to a ground state with ordered backbones as shown in Figure SI. 4.3b. The ordering of the backbones is indicated by the evolution of the average order parameter $\langle \Psi \rangle$, which is plotted in Figure SI. 4.4. The system energy per spin is also plotted in Figure SI. 4.4 in units $\epsilon_\parallel = 1$. The black curve shows the system energy at the end of each of the 500 temperature steps while the red curve shows an averaged value. Note that the ordering is not accompanied by a reduction in the energy, which in fact increases slightly because of thermal fluctuations.

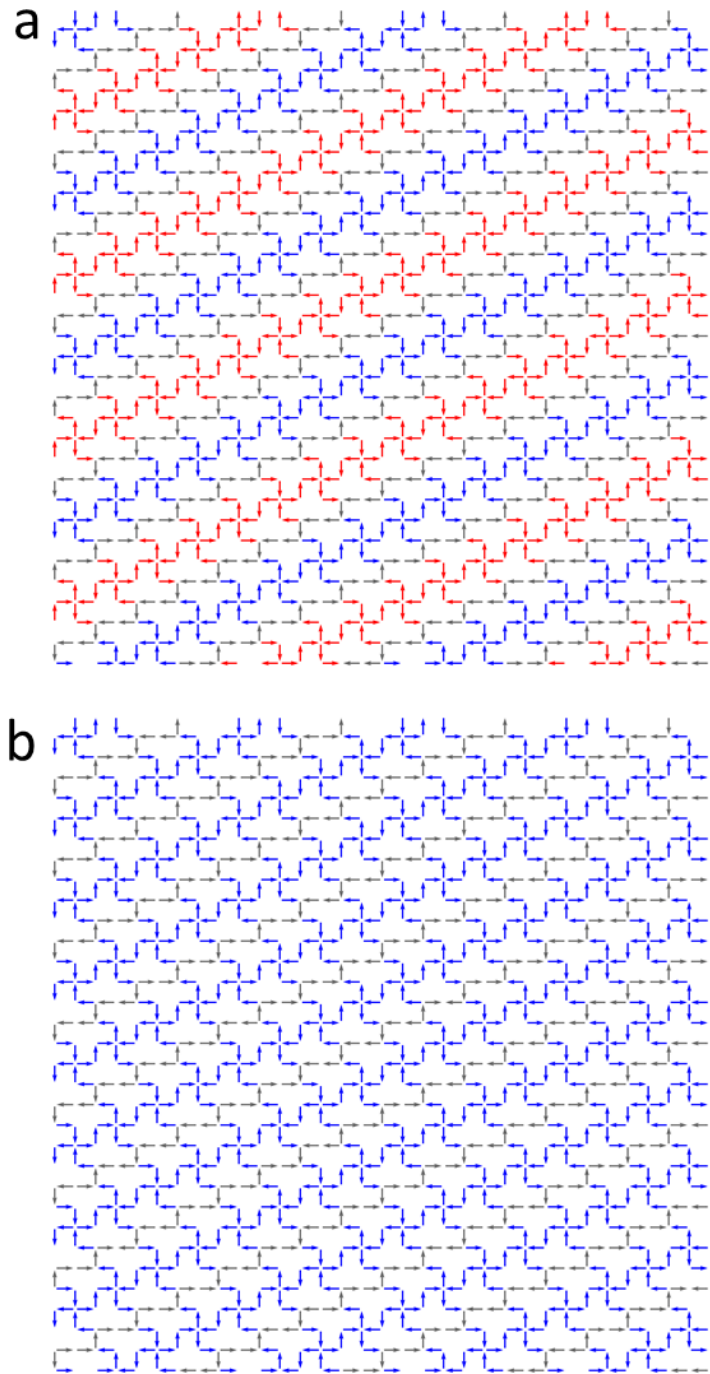

**Figure SI. 4.3:** (a) Simulation snapshot from the initial state where backbones alternate their order parameters. (b) Simulation snapshot from the final state with single domain ordering of the backbones. The color convention is identical to the experimental snapshots: blue and red arrows denote the backbone islands with opposite antiferromagnetic ordering, and gray arrows are the staircase moments. During the simulation, we kept the temperature at $k_BT/\epsilon_\parallel = 0.4$.

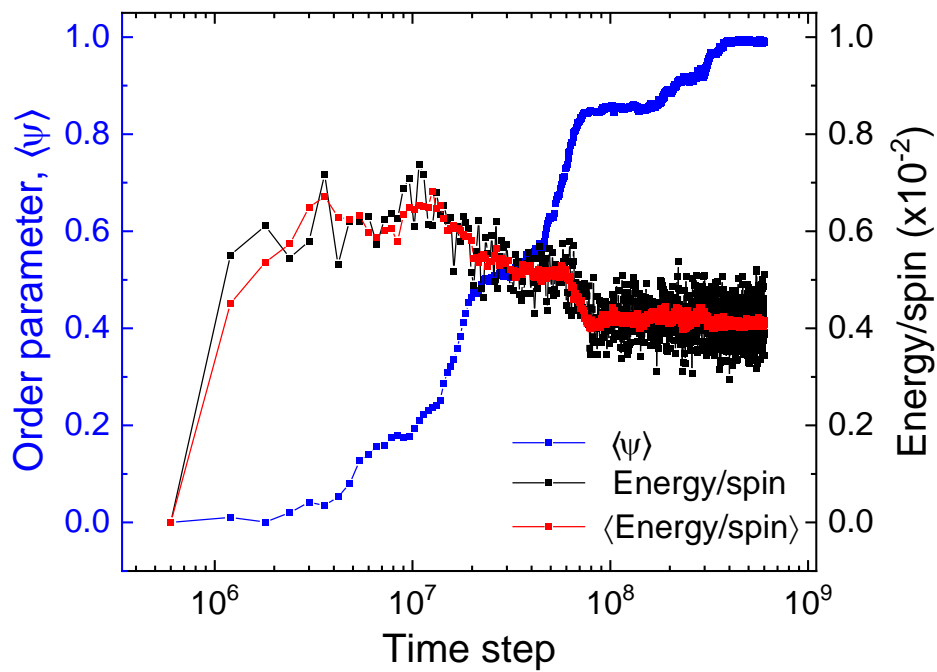

**Figure SI. 4.4:** Simulated order parameter as a function of Monte Carlo time step, starting from a state of alternating backbone moment order and evolving to uniform (single domain) backbone order. The red curve shows the average over the last $10^6$ time steps at that temperature. We initiated the system in a configuration, in which backbones alternated their order parameter $\langle \Psi \rangle$ and staircases were ordered with head-to-tail alignment, and the energy of this configuration was defined to be zero. The initial and final states of this simulation are shown in Figure SI. 4.3a and 4.3b, respectively.

## 5. Comparison of tetris ice with other models of entropy-induced ordering

Long-range ordering resulting from entropy considerations has been explored in different systems, often using slightly different terminology. The phrase "entropy-driven order" is generally employed for classical systems of driven particles, e.g., hard rods or colloids, where ordering of some degrees of freedom enables more entropy associated with other degrees of freedom (often in the spatial location of the constituent elements). By contrast, for spin systems, the phrasing "order-by-disorder" is more common to describe ordering that is driven by an optimization of free energy associated with thermal excitations among the spins, rather than by a reduction in energy; in these systems, the ground state is disordered, but thermal fluctuations select an ordered configuration. There have been a number of theoretical approaches to this issue, but a relative scarcity of experimental realizations. Tetris ice shares similarities with both the former (and in particular Onsager's model of thin rods [3]), and the latter (Villain's model of a two-dimensional spin system [4]).

Onsager's model of thin rods [3] exhibits nematic order because by lowering their orientational entropy, the rods attain higher translational entropy (rods can move around more easily if they are aligned). Though the two systems are completely different, the conceptual similitude with tetris ice is strong. In both systems, two ingredients lead to entropy-driven order. The first shared ingredient is the entropy, which in tetris ice is present even at very low temperatures because of the frustration. The second shared ingredient is a distinction between two kinds of entropies, so that the total entropy can increase at the expense of the reduction of one specific entropy, leading to partial order. In Onsager's model we have translational vs. nematic entropy. In tetris ice, we have the entropy of the staircase moments vs. that of the backbone moments.

Villain's study of a "domino system" bears *structural* resemblance to ours, but only to a degree. In Villain's model, two kinds of alternating one-dimensional Ising systems (called A and B-chains) are longitudinally ordered. The A-chains are ferromagnetic, and the B-chains are antiferromagnetic. A- and B-chains are coupled but there is no coupling among A-chains (much like there is no significant coupling among our backbones). In the ground state, there is no mutual ordering among A-chains because the antiferromagnetic order of the B-chain mediates no entropic interaction among A chains at zero temperature. However, at nonzero temperature, the A-chains become mutually ordered, because fluctuations in the B-chains have lower energy when A-chains order.

In tetris ice, the entropy-driven order is arguably more robust and significant than in Villain's model. Indeed, it is already present in the ground state, not only despite, but because of tetris' residual entropy. Conversely, Villain's model possesses no residual entropy, the disorder of its ground state scaling sub-extensively. There, calling $L$ the number of A-chains, the entropy of the ground state is $S_{Villain} = L \ln 2$, subextensive in the size of the system. The probability $p_{ORD}$ that a random configuration of minimal energy shows transverse order among the A-chains scales with $L$ as $p_{ORD} \propto 2^{-L}$ and tends exponentially to zero in the limit of a large system: most configurations in the ground state of Villain's model have mutually disordered A-chain orientation. The opposite is true for tetris ice. There, the probability that a randomly chosen configuration in the ground state shows full ordering among the backbones tends to approach 1 in the large system limit. This is due to the fact that staircases, unlike B-chains in the Villain model, can possess a linear density of entropy.

To show this, consider again a finite, square portion of the tetris system that contains $L$ backbones and $L$ staircases. We call $p_n$ the probability that a randomly chosen ground state configuration possesses $n$ disordered staircases (that is, staircases sandwiched among backbones of the same mutual orientation). The number of ways it can be realized is proportional to $\binom{L}{n} w^{nL}$, where $w$ is a number larger than 1 (because it can be written as $w = \exp(c s_{sc})$, where $s_{sc}$ is the residual entropy per spin of a disordered staircase and $c$ is some positive constant related to the linear density of moments) while $\binom{L}{n}$ counts the degeneracy in assigning the disordered staircases. Thus, by normalization, we obtain the probability of a ground state configuration with $n$ disordered staircases as $p_n = \binom{L}{n} w^{nL}/(1+w^L)^L$, and we see that $p_n$ has a maximum for $n = L$, corresponding to all staircases being disordered, and therefore all backbones being mutually aligned.

We have therefore that the probability of an ordered state, $p_{ORD}$, is

$$p_{ORD} = p_L = \frac{w^{L^2}}{(1+w^L)^L} \tag{5.1}$$

We see therefore that $p_{ORD} \to 1$ for large $L$, which is the thermodynamic limit. Note that $p_{ORD}$ would instead tend to zero if the staircases had no entropy $s_{sc} = 0$ leading to $w = 1$, and returning the same $p_{ORD}$ of Villain's model, as discussed above.

Therefore, when randomly picking a configuration in the ground state, the probability that it is backbone-ordered tends to one as the size of the system grows. This differs substantially from Villain's model, where, instead, thermal fluctuations must be invoked to induce order. It is, in fact *conceptually* more similar to the case of Onsager's model for rods, which applies to an athermal situation.

The argument can be framed entropically because, naturally, numbers of states are related to entropy. So the state of maximal probability, for which backbones are ordered, is also a state of maximal entropy. From the above, the entropy of a configuration with $n$ disordered backbones, $S_n$, is proportional to the logarithm of $\binom{L}{n} w^{nL}$, or $S_n \propto \ln\left[\binom{L}{n}\right] + nLc\, s_{sc}$ which is maximal for $n = L$ when $L$ is large enough.

The fact that the ordering in tetris ice comes from counting arguments in the ground state, without having to invoke the energy of thermal fluctuations, suggests that its entropy-driven order might manifest in a variety of scenarios, not necessarily at thermal equilibrium. Whatever mechanism produces ground state configurations in an unbiased way will eventually tend toward backbone ordering, because of the residual entropy. In that sense, tetris could be considered as an example of a more robust entropy-based ordering, rather than order induced by thermal fluctuations.

## 6. Explaining the vertical staircase moment "half ordered" correlations

From the Ising model for the staircases as reported in [2], once the horizontal staircase spins are chosen, the direction of vertical staircase spins is determined by those and by the connectors. Note that the connectors belong to backbones and are assumed to be ordered in the ground state both longitudinally and transversely. Interestingly, the correlations among vertical staircase spins are a mix of the correlations of the horizontal staircases and of the connectors because they inherit features of disorder from the former, and order from the later.

Consider Figure SI 6.1 as a reference for the symbols used below. As we prove later, for two vertical staircase spins $\sigma_i, \sigma_k$ in the same staircase, the following *exact* formula applies:

$$\langle \sigma_i \sigma_k \rangle = \frac{1}{16} \langle v_i v_k \rangle (9 - 6\langle S_i S_{i+1} \rangle + \langle S_i S_{i+1} S_k S_{k+1} \rangle) + \frac{1}{16}(2\langle S_i S_k \rangle + \langle S_{i+1} S_k \rangle + \langle S_i S_{k+1} \rangle)$$

(6.1)

In the limit of distances $k - i$ bigger than the horizontal staircases' correlation length, we have $\langle S_i S_k \rangle \sim 0$. Because in the ground state $\langle v_i v_k \rangle = (-1)^{|k-i|}$, the formula reduces to the sign-alternating

$$\langle \sigma_i \sigma_k \rangle \approx \left(\frac{9}{16} - \frac{6}{16} \langle S_i S_{i+1} \rangle\right)(-1)^{|k-i|} \tag{6.2}$$

where $\langle S_i S_{i+1} \rangle \sim 0.17$ in a vertex model, while $\langle S_i S_{i+1} \rangle \sim 0.4$ experimentally from previous work [2], as we also show in Figure 4b in the main text.

**Longitudinal Correlations:** Consider first the pure vertex-model, simulated by our Monte Carlo algorithm. We know that spin correlations are very small, or $\langle S_i S_k \rangle \approx (0.172)^{|k-j|}$, and therefore the first term of Eq. (6.1), or

$$\langle \sigma_i \sigma_k \rangle \approx \left(\frac{9}{16} - \frac{6}{16} \langle S_i S_{i+1} \rangle\right) \langle v_i v_k \rangle \tag{6.3}$$

dominates. Then in the ground state we know that connectors alternate and thus $\langle \sigma_i \sigma_k \rangle \approx 0.499(-1)^{|k-1|}$, explaining the constant alternation seen in our simulations. Above the ordering transition, instead, $\langle v_i v_k \rangle \approx 0$ and the alternating correlations among vertical staircase spins disappear because connectors are no longer ordered, as also seen in our simulations (Figure SI. 4.2e and 4.2f).

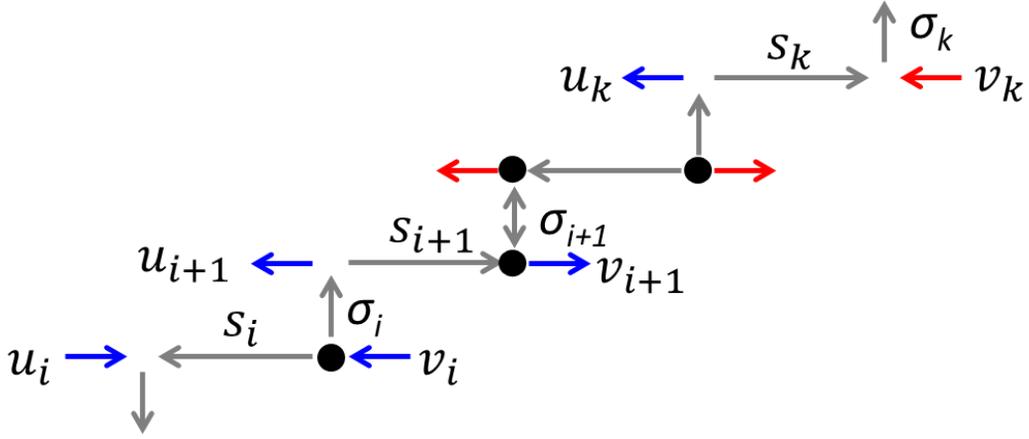

**Figure SI. 6.1** Schematic of spin labeling used in this section. Spins pointing leftward (rightward) and upward (downward) were defined to be positive (negative). For instance, $S_i = 1$, $u_i = -1$, $v_i = 1$, and $\sigma_i = 1$. Blue and red correspond to the backbone spins with opposite antiferromagnetic ordering, and gray represents the staircase spins. Black dots are the excitations.

In the case of the real material, $\langle S_i S_k \rangle$ is in generally larger, for reasons explained and computed in the previous publication [2], and as we see in Figure 4b in the main text. However, we have in general $\langle \sigma_i \sigma_k \rangle \leq \langle \sigma_i \sigma_{i+1} \rangle \approx 0.4$. Therefore, all the terms except the first, alternating one, in Eq. (6.1), contribute less than 0.1. Nonetheless, they show up in a slight dependence on the distance $|k - i|$ that can be detected in the data. Then, in the ground state, at large distance $|k - i|$, where the horizontal staircase spin correlations are zero, the alternating correlation of Eq. (6.2) survives, or $\langle \sigma_i \sigma_k \rangle \approx 0.41(-1)^{|k-1|}$.

**Transverse Correlations:** From the way it is deduced (see below), Eq. (6.1) also applies to the transverse correlations, for moments defined as in Figure SI. 6.1, when now $i, k$ denote not the position of the vertical staircase spins within the staircase, but rather numbers the vertical staircase itself transversely.

There too the same considerations above apply. Except now the alternation of connectors $\langle v_i v_k \rangle = (-1)^{|k-i|}$ is due to transversal ordering of the backbones.

**Proof of the equation:** Starting from $\langle \sigma_i \sigma_k \rangle$ the crucial point is to express $\sigma_i$ in terms of nearby horizontal staircases $S_i$ and connectors $u_i, v_i$ as illustrated in Figure SI. 6.1. We will consider spins positive if they point left or up (any convention is equivalent).

Note that when both $S_i, v_i$ and $S_{i+1}, u_{i+1}$ are antiferromagnetic, then $\sigma_i$ has the same sign of $v_i$. If they are ferromagnetic and $S_{i+1}, u_{i+1}$ are antiferromagnetic, then $\sigma_i$ has the same sign of $u_{i+1}$. If they both $S_i, v_i$ and $S_{i+1}, u_{i+1}$ are ferromagnetic, this means that the vertical staircase spin $\sigma_i$ has two unhappy vertices at its' ends and it is thus a stochastic variable $\sigma_i = \theta_i$ equal to +1 or -1 with equal probability. We can therefore write

$$\sigma_i = \frac{1-v_iS_i}{2}v_i + \frac{1+v_iS_i}{2}\frac{1-u_{i+1}S_{i+1}}{2}u_{i+1} + \theta_i \qquad (6.4)$$

where $\theta_i$ is different from zero if and only if both the first two terms are zero, thus implying an unhappy vertex at both ends of the vertical staircase spin. Now, assuming we are in the ground state, using $v_i = u_{i+1}$ and $u^2 = v^2 = 1$ we obtain from Eq. (6.4)

$$\sigma_i = \frac{3}{4}v_i - \frac{S_i+S_{i+1}}{4} - v_i\frac{S_iS_{i+1}}{4} + \theta_i \qquad (6.5)$$

Plugging Eq. (6.4) into $\langle \sigma_i \sigma_k \rangle$ and remembering that: a) free vertical staircase spins are uncorrelated and therefore $\langle \theta_i \theta_k \rangle = \langle \theta_i \rangle = 0$, b) horizontal staircases are disordered and therefore $\langle S_i \rangle = 0$, we obtain general formula Eq. (6.1). Naturally, if we are not in the ground state, its expression is altered by additional terms associated with the presence of excitations among the moments.

**Supplementary Information References**